\documentclass[sensors,article,accept,moreauthors,pdftex,obeyspaces]{Definitions/mdpi} 

\graphicspath{images/}
\firstpage{1} 
\pubvolume{1}
\issuenum{1}
\articlenumber{0}
\pubyear{2022}
\copyrightyear{2022}
\externaleditor{Academic Editor: Yuh-Shyan Chen 
} 
\datereceived{23 December 2021} 
\dateaccepted{24 January 2022} 
\datepublished{} 
\hreflink{https://doi.org/} 

\usepackage[normalem]{ulem} 

\usepackage{tikz} 

\Title{Ransomware: Analysing The Impact on Windows Active Directory Domain Services}

\TitleCitation{Ransomware: Analysing The Impact on Windows Active Directory Domain Services}

\Author{Grant McDonald, 
 Pavlos Papadopoulos *\orcidA{}, Nikolaos Pitropakis *\orcidB{}, Jawad Ahmad \orcidC{} and~William J. Buchanan~\orcidD{}}

\AuthorNames{Grant McDonald, Pavlos Papadopoulos, Nikolaos Pitropakis, Jawad Ahmad and William J. Buchanan}

\AuthorCitation{McDonald, G.; Papadopoulos, P.; Pitropakis, N.; Ahmad, J.; Buchanan, W.J.}

\address[1]{%
Blockpass ID Lab, School of Computing, 
Edinburgh Napier University, Edinburgh EH10 5DT, UK; grant.a.mcdonald@gmail.com (G.M.); j.ahmad@napier.ac.uk (J.A.); b.buchanan@napier.ac.uk (W.J.B.)} 

\corres{\hangafter=1 \hangindent=1.05em \hspace{-0.82em}Correspondence: pavlos.papadopoulos@napier.ac.uk (P.P.); n.pitropakis@napier.ac.uk (N.P.)}

\abstract{{Ransomware has become an increasingly popular type of malware across the past decade and continues to rise in popularity due to its high profitability. Organisations and enterprises have become prime targets for ransomware as they are more likely to succumb to ransom demands as part of operating expenses to counter the cost incurred from downtime. Despite the prevalence of ransomware as a threat towards organisations, there is very little information outlining how ransomware affects Windows Server environments, and particularly its proprietary domain services such as Active Directory. Hence, we aim to increase the cyber situational awareness of organisations and corporations that utilise these environments.} Dynamic analysis was performed using three ransomware variants to uncover how crypto-ransomware affects Windows Server-specific services and processes. Our work outlines the practical investigation undertaken as WannaCry, TeslaCrypt, and Jigsaw were acquired and tested against several domain services. The findings showed that none of the three variants stopped the processes and decidedly left all domain services untouched. However, although the services remained operational, they became uniquely dysfunctional as ransomware encrypted the files pertaining to those services.}


\begin{document}


\section{Introduction}


There is no questioning that information technology (IT) and computing play an integral part in the day-to-day operations of enterprises and organisations in modern society. IT systems have immeasurably increased productivity in the modern workplace, and as a result, a~dependency upon this has been created, so much so that ``IT services are becoming a critical infrastructure, much like roads, electricity, tap water, and~financial services'' \cite{Franke2017}. When IT systems stop functioning in business environments, companies can lose a large amount of money through non-utilised staff wages, missed opportunities, and~reputational harm, with~the average cost of downtime totalling \$141,000~\cite{DattoInc.2020}. Cybercriminals have caught on to this and have begun to take advantage of the harm caused by data destruction and downtime by using a particular form of malware called ransomware. Designed to hold the system or its contents hostage until a ransom is paid, they are particularly damaging to organisations due to the aforementioned consequences of downtime, making organisations much more lucrative targets. The~profitability of ransomware relies upon the willingness to pay the ransom, and~when the cost of downtime is 23 times greater than the average ransom demand of USD 5900, it is no surprise that the ransomware industry continues to grow~\cite{DattoInc.2020}.

With downtime having the largest financial impact when it comes to corporate IT utilisation, in~conjunction with the threat of blackmail from stolen files, succumbing to the ransom demand becomes very appealing. In~a 2018 study, researchers were able to trace an estimated USD 16 million in ransom payments through a two-year period from a potential 19,750 victims~\cite{Huang2018}, with~a further estimated total of over USD 25 million in payments between the years 2016 and 2017~\cite{Ramsey2017}. The~ransomware SamSam alone had netted its developers USD 6.5 million over the course of just under 2 years~\cite{Sophos2019}, with~its highest single ransom payment recorded at USD 64,000~\cite{Sophos2018}. Although~ransomware profits seem exorbitant, the~cost of damages is even more astounding. Deep Instinct estimates that the total damage cost of ransomware in 2019 exceeded the predicted USD 11.5 billion, as~well as stating that ransomware developers specifically targeted large enterprises due to their profitability~\cite{DeepInstinct2020}. As~such, extra precautions should be taken by organisations and enterprises to ensure they reduce their likelihood of becoming a ransomware victim. One method of reducing this possibility is by creating a plan based upon all the information available, and~where there is a lack of information available, the~gap must be~filled.

{Typically, exploiting less technologically skilled users would be the easiest pathway into a network, as most staff at companies outside of the IT industry will only learn the IT skills required to perform their job effectively. These IT skills, in~the eyes of the user, would not include being able to effectively evaluate and put a stop to any potential cyber threats. Therefore, domain controllers that are ideally only operated by trained cyber situational aware IT professionals should theoretically be less susceptible to threats than devices operated on by those less technologically skilled. However, as~malware continues to evolve, threat vectors are shifting. WannaCry, one of the most notorious ransomware variants seen in recent history, has the ability to spread across hosts on a network without user interaction by exploiting a network protocol vulnerability and is by no means the only form of ransomware able to do so. As~such, domain controllers operating on Windows Server, which face increased network exposure to offer their services, are just as susceptible to modern ransomware as regular consumer Windows versions. Despite there being a presence of ransomware-related queries regarding Windows Server, and~furthermore, domain controllers, posted by IT professionals on various internet forums and discussion boards, there appears to be a distinct lack of academic material and information regarding this specific topic. Therefore, it lies on our work to shed light on this case and investigate whether ransomware can impact the functionality of Windows Server domain services, and~to what extent.}

\subsection*{Contributions}

\begin{figure}[H]
\includegraphics[width=0.8\linewidth]{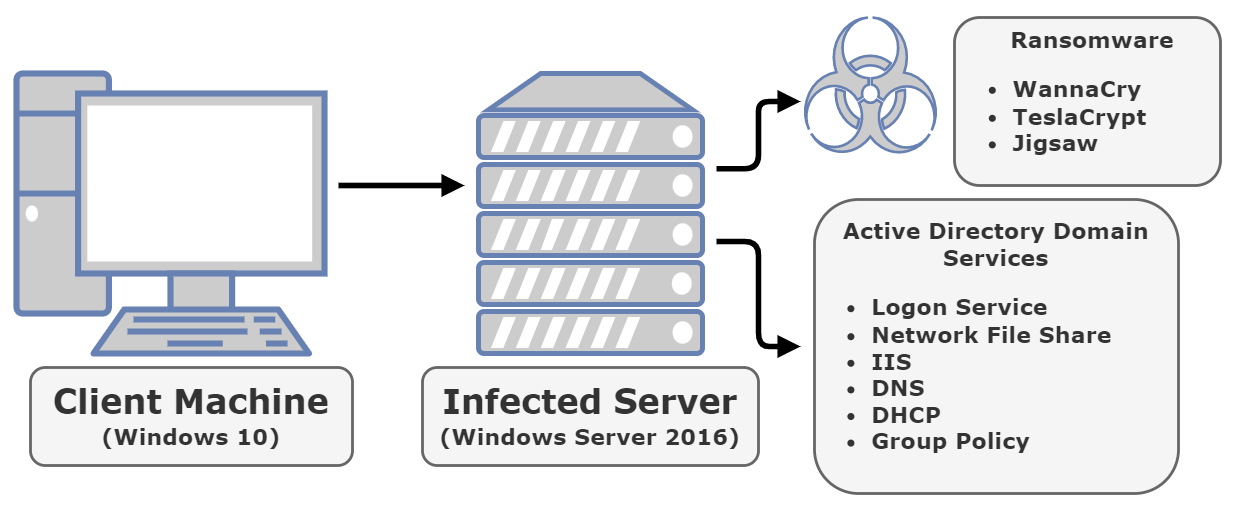}
\caption{{Overview of the investigation~approach. \label{fig:architecture}}}
\end{figure}

The overview of our investigation approach is illustrated in Figure~\ref{fig:architecture}. Additionally, the~contributions of our work can be summarised as follows:

\begin{itemize}
    \item We conceptualise and create an up-to-date test bed environment after extensively examining the literature regarding the functionality of Windows Server and Active Directory Domain Services;
    \item We launch three different ransomware attacks against the test bed environment while thoroughly presenting the results of the introduced experiments;
    \item Finally, we present the~analysis and critical evaluation of the experimental findings, unveiling their importance for modern Active Directory infrastructures.
\end{itemize}


\vspace{-6pt}

Following this introductory section, the~structure of our work is as follows.
Section~\ref{sec:litreview} provides the literature review conducted, which involved researching 
Active Directory Domain Services, ransomware, and~relevant malware analysis tools.
Utilising the information gathered from the previous section, our design and methodology is presented in Section~\ref{sec:methodology}.
{Our experimental implementation is outlined in Section~\ref{sec:implementation}, which includes the ransomware execution, the~software utilised, and~the virtual machines and Windows Server services configuration.} 
The results from the implementation are presented, analysed and evaluated in Section~\ref{sec:evaluation}. 
Finally, Section~\ref{sec:conclusion} draws the conclusions while giving some pointers to promote future work that builds upon or improves the current~study.


 
\section{Literature~Review}
\label{sec:litreview}

This literature review will provide context relevant to the background of ransomware and its impact on enterprises. Aiming to demonstrate the relevance and importance of the work to be undertaken, this will be done by exploring ransomware history and mechanics, enterprise infrastructure, and~related~work.

\subsection{Active Directory and Its Enterprise~Adoption}

Active Directory is Microsoft's user directory service and is arguably the most popular solution for organisations to manage and organise their staff's IT profiles for authentication, authorisation, and~accounting purposes. The~level of popularity can be evidenced by the fact that 95\% of Fortune 1000 companies utilised the service in 2014~\cite{HelpNetSecurity2014}. Active Directory works on a network domain structure, and~as such, a~machine running Windows Server 2000 or newer is required to act as the domain controller to run the service. In~this context, a~domain can be defined as a ``distinct subset of the internet with addresses sharing a common suffix or under the control of a particular organisation or individual” \cite{Lexico2020}.
Active Directory offers several services useful for managing the IT infrastructure of an organisation. Active Directory's primary purpose is to provide authorisation, authentication, and~accounting measures to organisations for use by systems/network administrators. User profiles are required for users to log in to domain-connected machines. Once a user has signed onto a machine via authentication, their actions will be restricted based on authorisation and~logged through accounting. From~these user accounts, policies can be applied through Group Policy Objects (GPOs) for various workplace purposes, such as assigning user groups based on department, then assigning printer or file sharing to those users, or~any policy that the organisation requires. For~many organisations, these services are critical to business~operations.


Domain controllers are a critical component of most modern corporate network structures, thereby making downtime of these hosts unfavourable even to perform necessary tasks such as software updates to patch security flaws. Organisations are even further discouraged from updating to the newest major operating system version, such as from Server 2008 to Server 2012, due to the differences and incompatibilities in User Interface and service functionality, as~well as even further increased downtime. Microsoft typically supports their latest operating systems for up to 10 years following release~\cite{Microsoft2020a}. This support encompasses new features, improvements, bug fixes, and~most importantly, patching security vulnerabilities. After~10 years from the extended support plan, Microsoft will no longer provide security updates to their operating systems despite the possibility, and~inevitability, of~security vulnerabilities becoming newly discovered after this time. 
Notwithstanding the operating system's vulnerability, there are countless organisations that either neglect or refuse to upgrade their operating system to the latest major version. In~2019, Microsoft estimated that around 60\% of Windows Server installations were version 2008, which amounts to roughly 24 million Windows Server 2008 servers~\cite{Alspach2019,Sayer2020}, an~alarmingly high number of servers that were less than a year away from end of support. Alongside neglecting software updates, organisations often fail to allocate sufficient resources towards IT infrastructure such as a backup domain controller, which would be immensely advantageous in recovering from a ransomware attack. Smaller, lesser-employee and lower-revenue managed companies may even completely lack IT staff altogether, leaving no one capable of maintaining a domain controller. As~domain controllers offer various network-based services, they leave many vulnerabilities exposed. All these factors result in domain controllers being a pillar of any organisation's IT structure that, when crippled, will have a large impact on the functioning of the remaining hosts on the~network.

\subsection{Ransomware}

Although software can be created for any desired purpose, from~entertainment to aiding productivity in the workplace, it can also be used for nefarious purposes. Malware is defined as “software that is specifically designed to disrupt, damage, or~gain unauthorised access to a computer system” \cite{Lexico2020a}. Ransomware is a subset of malware designed to digitally extort its victims into paying a demanded ransom amount, and~it does so through two main methods~\cite{Liska2016}. The~first ransomware type, known as crypto-ransomware, encrypts user files whilst leaving the machine otherwise operational. Paying the ransom will then, in~most cases, return the decryption key to decrypt the user's files. The~other main form of ransomware, often called locker ransomware, does not encrypt the user's files and instead locks the user out of their device to prevent them from using it until the ransom is paid~\cite{Kaspersky2020}. 
Malware is generally very dependent on its host's networking capabilities, and~from botnets to spyware, ransomware is no exception. For~the victim to pay off the ransom, an~internet connection is required; therefore, the ransomware must leave networking capabilities functional or instruct the user to pay the ransom using another device. The~latter is not a preferable solution as the attacker would ideally aim to infect as many devices as possible, leaving no other devices free to do anything other than pay the ransom. Additionally, ransomware can spread across a network to infect additional hosts, furthering the damage or ransom potential to the attacker. This networking aspect is particularly damaging to companies that utilise and depend on internal domain structure networks with a large number of hosts for employee usage.
The crypto-ransomware's execution process has several stages, as~follows:

\begin{enumerate}
\item	The first stage involves the ransomware utilising an attack vector to find its way onto a potential victim's machine. Attack vectors can include email phishing, injection through compromised or outdated legitimate software, or~exploiting networking protocols, with~some ransomware variants making use of several methods to achieve success; 
\item	Once the ransomware has gained access to the system, it begins performing an initial execution phase. At~this point, the~malware buries itself into the system and implements persistence mechanisms to automatically restart upon events such as system reboots~\cite{Brewer2016}; 
\item	The next action that the ransomware executable takes is to remove all backups that the user could utilise to circumvent the ransom payment. In~Windows, a~common method often used by ransomware, including WannaCry, is through the Volume Shadow Copy service~\cite{Adamov2017};
\item   Following this, the~ransomware will begin its encryption process. This process is also not uniform, as encryption algorithms and methods will differ across ransomware variants. Most ransomware variants use symmetric private key encryption, commonly AES, for~encrypting the file system with an asymmetric public key pair, commonly RSA, from~the Command and Control (C\&C) server to encrypt the randomly generated private key. WannaCry makes use of an AES and RSA combination~\cite{Subedi2018}. However, this is not the case with some ransomware variants, such as SamSam, that do not communicate with a C\&C server and instead complete the encryption phase locally, as the attacker has remote control of the machine. The~encryption component is the most distinguishable feature of each ransomware family, alongside the amount of ransom demanded~\cite{Hassan2019}. This is also obviously the most devastating component of crypto-ransomware, as once the encryption process has begun, the~ransomware process has become almost irreversible. While some ransomware variants do have their decryption tools made freely and publicly available through work conducted by researchers, they have usually only been made so after the initial outbreak, when organisations have already made decisive action on how to proceed with their infected systems; 
\item   The final event of the ransomware execution process is the ransom demand notification. This involves prompting the user with a notification window stating that they have been infected. This window will outline the ransom amount required and instructions on how to pay, and~in some instances also includes a countdown that, when expired, will either increase ransom demand and restart the countdown, or~wipe the system.
\end{enumerate}


{While other malware such as spyware, botnets, and~rootkits thrive on remaining undetected to the user, ransomware is the opposite. Ransomware's capability of “incapacitating the core business functions of a system” \cite{Zimba2019} is what makes it particularly damaging and unpleasant, resulting in enterprises becoming much more willing to take action, including paying the ransom, to~return functionality to normal. Another contributing factor towards a company's willingness to pay a ransom is reputation. While becoming the victim of a cyberattack can damage the reputation of the organisation, the~reputation of the ransomware's credibility is equally important towards the functioning of the ransomware ecosystem. Another critical point towards understanding the ransomware ecosystem is that the “ability to make money from ransomware critically depends on victims believing that the criminal will honour ransom payments.” \cite{Cartwright2019}. In both Q3 and Q4 of 2019, 98\% of the companies that paid the ransom were supplied with a working decryption tool, proving that ransomware developers are set on keeping the trust~\cite{Coveware2020a}. The~developers of the ransomware variant SamSam went as far as offering technical support to their victims to ensure that their data was recovered following payment~\cite{Sophos2018}. The developers of UltraCrypter also took this approach after it was discovered that their payment system seemed to be dysfunctional~\cite{Abrams2016a}. However, perhaps the most influential incentive to pay the ransom is cybersecurity insurance. Of~the organisations that paid the ransom, 94\% were reimbursed through their insurance~\cite{Sophos2020}.}

\subsubsection{{Ransomware-as-a-Service}}

An increasing trend in the ransomware cybersphere is the use of Ransomware-as-a-Service (RaaS). RaaS allows those without technical malware development skills to partake in the activity of deploying and utilising ransomware by using another ransomware developer's creation to do so. Ransomware developers may offer their code outright for a set price, or~simply supply the ransomware gratuitously under the condition that the perpetrator shares a percentage of the profits with the developer. While it may seem counterproductive for ransomware developers to let others use their creation and partially reap the resultant profits as opposed to performing the act themselves, there are several reasons for this industry's existence. By~offering RaaS, developers reduce their risk of exposure, as they sell their product anonymously through the darknet and only leave a financial trail through almost untraceable cryptocurrency so that not even the end customer can identify the developer if caught. Furthering this, regular individuals have access to areas that malware developers may not. For~example, a~disgruntled employee may use their corporate access to deploy ransomware to domains and networks that would have otherwise been difficult or impossible for the initial developer to penetrate~\cite{Meland2020}. This has opened yet another channel for ransomware to thrive in the modern cyber~landscape. 


As ransomware continues to evolve, so do the strategies that developers employ to gain larger profits. Ransomware developers “will slowly discover and converge towards optimal strategies” to produce more desirable results~\cite{Hernandez-Castro2017}. This can already be seen from several examples used in modern-day ransomware variants. A~new common strategy employed to effectively target enterprises is the threat of leaking, or~making publicly available, sensitive files acquired during the ransomware infection. This amounts to stealing sensitive files such as patents from companies, or~personal information from targets like hospitals, and~sending them back to the ransomware perpetrator to further extort and encourage payment~\cite{NationalCyberSecurityCentre2020}. Although~a domain controller would typically not be used to work on sensitive documents, they may still be used to store them in the form of a network file share. Therefore, the~network file share server would be a lucrative target, as it is an aggregate of the work of several users, as~opposed to gaining access to a host belonging to and storing the work of one~user.

%




\subsubsection{{The Role of Cryptocurrencies in the Ransomware~Industry}}

While 
 crypto-ransomware relies on encryption to perform its function, the~developer depends upon cryptocurrency to reap its profits. Cryptocurrency is a term used to describe digital currency where cryptography is used to verify transaction records and ledger ownership. The~role of cryptocurrency in ransomware is that it provides the attacker with an almost anonymous financial account to receive the profits of their attacks without leaving a clear financial trail to their real, physical identity. This has made deploying ransomware a very low risk, potentially high reward cybercrime. Bitcoin is undoubtedly the most well-known, as it was the first decentralised digital currency and has ultimately become one of the highest valued~\cite{papadopoulos2021decentralised}. While Bitcoin can be considered anonymous in the sense that no information regarding the account represents real-world identifiable information that can be tied to an individual, the~account is still represented by an address that, as~a blockchain-based currency, can be easily attributed and linked publicly to all of its transactions. As~a result, Bitcoin is not entirely anonymous and is instead labelled pseudonymous~\cite{Bistarelli2018}. 

When exchanging to real currency, it would be possible for an investigator to use the payment from a ransomware-operated ledger to an exchange, then match the respective value from the exchange to a suspect's real bank account. Furthermore, cryptocurrency exchanges are often real businesses that are regulated by local authorities and would therefore be subject to any local laws that could force the exchanges into providing information regarding transactions to real bank accounts, ultimately uncovering the real identity of a cybercriminal~\cite{Kshetri2017}. Despite attackers being able to use multiple Bitcoin accounts to split up and effectively launder the currency, the~transaction trail will always remain public and traceable due to Bitcoin's blockchain infrastructure. {However, in~case a cybercriminal uses a centralised mixer to exchange Bitcoin to a more privacy-focused cryptocurrency such as Monero, and~repeat this process through multiple different exchanges, tracing of the transactions becomes impossible to follow, as~these cryptocurrencies obfuscate transaction records, thereby removing the public transaction trail. Utilising multiple mixer exchanges across various law enforcement jurisdictions then makes tracing transactions much more legally complex to undertake. With~all these factors, it is no surprise that the proliferation of ransomware continues year by year~\cite{young2021evaluating}.}

\subsection{TeslaCrypt and~Jigsaw}

{Two notable ransomware variants that are used for the practical experiment are TeslaCrypt and Jigsaw. TeslaCrypt's existence was first identified in early 2015 and was reportedly spread through spam mailing lists and compromised websites~\cite{Lemmou2018}. It was initially designed to target data belonging to video games, including save files and profiles; however, at~some point it was altered by the developers to include a wider file range, possibly to increase profitability through a wider range of victims. TeslaCrypt demanded a ransom of USD 500 equivalent in Bitcoin, which would double every 60 h~\cite{Adamov2017}. AES-256 encryption was used by TeslaCrypt; however, due to a bug introduced in the first TeslaCrypt iteration, the~encryption process was reversible. This was resolved by version 2, and the ransomware remained so through the following versions until the campaign came to an end. The~TeslaCrypt campaign came to an end when, in~May 2016, the~developers released the master decryption key on their Tor-hosted payment website~\cite{Mimoso2016}. This allowed for those infected to decrypt their files, and~for software developers to release decryption tools. Although~the total statistics of TeslaCrypt's financial impact are unknown, between 7~February and 28 April 2015, the developers gained a sum of \$76,522 through PayPal and Bitcoin payments from 163 victims~\cite{Villeneuve2015}. In~comparison to more notorious ransomware variants, TeslaCrypt seems unimportant to the crypto-ransomware scene; however, this still has a large impact on the victims affected. The~lack of notoriety arguably makes the ransomware more impactful, as it attracts less attention from malware analysts that would work towards a solution.}

{Jigsaw is a relatively unknown ransomware variant that did not gather mainstream popularity in the same way as other ransomware variants did.} Jigsaw received its name from its depiction of a character in the popular movie series ``Saw'' in its ransom note. The~horror movie depictions have also caused Jigsaw to be classified by some as scareware. The~ransom demanded by Jigsaw is either USD 150 equivalent in Bitcoin or 0.4 Bitcoin. To~encourage payment, the~ransomware variant claims to delete several files per hour, increasing in amount each time, until~either the ransom is paid, or~72 h have passed. Jigsaw was the first variant to introduce the gradual file deletion mechanism, and after 72 h had passed since execution, all files remaining would be deleted~\cite{OKane2018}. On~top of this, Jigsaw would delete 1000 files each time the computer is restarted. From~March to August 2016, a~total of 2.5 Bitcoin was reportedly sent from 58 victims to Jigsaw-operated Bitcoin addresses~\cite{Conti2018}.

\subsection{WannaCry}

On 12 May 2017, the WannaCry outbreak began. The~ransomware gained significant media attention within hours, as it crippled several major institutions and critical infrastructure in Europe such as the United Kingdom's National Health Service, Deutsche Bahn, Renault, FedEx, and~several other high-profile organisations. WannaCry affected over 300,000 businesses across 150 countries in the first few days of its outbreak~\cite{Europol2017}. Many cybersecurity investigators and analysts, including The National Cyber Security Centre of the UK, had suggested that the Democratic People's Republic of Korea (North Korea) were behind the attack due to similarities in previous attacks that were also attributed to them~\cite{NationalCyberSecurityCentre2017}.
{To complete its task, WannaCry made use of two exploits developed by the USA's National Security Agency (NSA) that were leaked earlier in the year by a hacker group that called themselves ``The Shadow Brokers''. The~first exploit tool used is known as ``DoublePulsar'' and is a backdoor that allowed any unauthorised users to execute the malware on a machine without the user's interaction.} This was used in conjunction with another exploit known as ``EternalBlue''. EternalBlue is the name given to an exploit that allowed the ransomware to spread rapidly across a network. It does this by exploiting a vulnerability in the Server Message Block (SMB) protocol, a~protocol most commonly used by Microsoft for network file and printer sharing in Active Directory~\cite{Microsoft2018a}. Although~there are several updated versions of SMB, only version 1 contains the vulnerability and is still enabled by the Active Directory service to offer its services to older clients that may not support the newer protocol versions. Windows displays this file sharing service as ``LanmanServer'' which automatically starts upon booting Windows, including its support for SMB version 1~\cite{RevertService2020}. However, in~the most recent Windows installations, SMB version 1 is not installed by default and, for~older installations, Microsoft now recommends disabling version 1 support if possible~\cite{Microsoft2020,Microsoft2020b}. {As such, almost contradictorily, this exploit made the usage of Windows Active Directory a~large security vulnerability for organisations; thus, this was the~reason for WannaCry's large-scale targeting of enterprise information systems.} 

As noted previously, enterprises generally have not kept software updated as frequently as is required. WannaCry's outbreak demonstrated this greatly, as the vulnerability that EternalBlue exploited was patched in security update MS17-010, which was released almost 2 months prior to WannaCry's outbreak. Despite the MS17-010 update being labelled as critical, WannaCry took only hours to spread globally~\cite{Microsoft2017}. The~vulnerability was listed as CVE-2017-0144 and stated that Windows Server versions up to 2016 are all affected~\cite{TheMITRECorporation2017}. 
On the same day that WannaCry began to spread rapidly, a~malware analyst known as MalwareTech had begun to analyse a sample of the ransomware. They discovered that before WannaCry executes, it attempts to contact a domain which, at~first, was unregistered. MalwareTech then registered the domain under the assumption that they had taken control of the C\&C server when, in~reality, they had instead accidentally activated the ransomware's kill switch, as the ransomware would cease execution if it was able to contact the domain~\cite{MalwareTech2017}. This resulted in newer infections ceasing to execute, whilst devices that had already passed WannaCry's kill switch check would remain infected and encrypted. Although~WannaCry would not begin to execute on devices with internet access, it could continue to spread and activate on devices that did not have an internet connection, as it would fail to contact the kill switch domain. 
Domain controllers typically only need to provide services to those on its domain, and~in some instances, this meant that organisations would, as~a security precaution, revoke public internet access from the domain controllers. In~the case of WannaCry, a~decryption tool was made available a week after its initial outbreak, at~which point the kill switch had already been activated to thwart new infections, and~previously infected users had already taken action by either paying ransoms, restoring from backups, or~wiping infected machines and restoring from scratch~\cite{Suiche2017a}.









\subsubsection{The Impact of WannaCry and Current~Trends}

WannaCry's impact and subsequent news coverage brought the topic of cybersecurity into the mainstream. The~attack forced organisations outside of the IT industry to evaluate their own security measures and consider whether they should improve their IT infrastructure. Its mainstream popularity even had implications unrelated to cybersecurity practice, as the finance sector had capitalised upon the event and saw excess positive returns in cybersecurity exchange-traded funds as a result of WannaCry~\cite{Castillo2018}. 
Despite the publicised impact of ransomware at that point in time, it seems that enterprises have failed to learn the lesson from those previously affected. Exactly 2 years after the debut of WannaCry on \mbox{12 May 2017}, it was uncovered through the internet device search engine Shodan that over one million computers and servers still had SMB version 1 enabled~\cite{SentinelOne2019,Whittaker2019}. This number does not account for devices that were not public-facing, and~therefore does not include devices that Shodan could not locate. 
Even more worrying is that despite the NHS having been crippled for days as a result of WannaCry, it was also revealed at around the same period in 2019 that approximately 2300 NHS computers were still operating on Windows XP which, at~the time, was over 5 years past its end of support date~\cite{Doyle-Price2019}. While the ransomware attack did cause appointment cancellations and impacted hospital admissions, it thankfully did not impact the mortality rate. Financially, the~event caused an estimated loss of £5.9 million to the NHS through lost activity~\cite{Ghafur2019,stamatellis2020privacy}. 

However, the~story of WannaCry is yet to end. As~the kill switch has been exposed, the~ransomware will no longer run on devices with a public internet connection capable of contacting the hardcoded domain. In~network environments where that connection is not possible, the~ransomware will fail the check to cease execution and will begin to freely infect the network once again. In~August of 2018, over~a year after the initial WannaCry outbreak and its subsequent kill switch exposure, the~Taiwan Semiconductor Manufacturing Company (TSMC) was infected by a WannaCry variant that affected more than 10,000 computers~\cite{SkyboxSecurity2018}. TSMC's internal network structure was set up with security in mind, which resulted in a structure that did not allow hosts on the network to have internet access. Ideally, this would massively reduce threat vectors, particularly the RDP vector discussed previously. However, this structure was the one flaw that denied their network protection from WannaCry. After~an individual connected a new device to the network that had previously obtained the WannaCry virus from the web, the~ransomware failed to locate the kill switch domain and immediately began to spread across TSMC's internal network without stopping due to the air-gapped network design. TSMC is not the only high profile enterprise to become a late victim to WannaCry, as~just months before in March of 2018, Boeing was also affected, albeit not to the same extent as TSMC or those from the initial outbreak~\cite{Goud2018,Muncaster2018}. 

WannaCry's popularity caused it to spawn many offshoots, as~people saw opportunity from its debatable success. In~2019, an~estimate suggested that there were more than 12,000~WannaCry variants with even just basic kill switch removal modifications~\cite{Mackenzie2019}. Recent events and ransomware variants have proven that despite the widespread publication of WannaCry's devastation, there is still an alarmingly high number of devices unpatched and~just as vulnerable to the same exploit used years~before.

\subsubsection{NotPetya}

One notable example of ransomware is the NotPetya variant, which was released in June 2017, shortly after WannaCry, and~even used the same EternalBlue exploit. However, the~purpose, execution, and~deployment method varied significantly from WannaCry. The~infection began as cybercriminals had managed to infiltrate the update server of Ukrainian accountancy software used by an estimated 80\% of companies in Ukraine~\cite{BritishBroadcastingCorporation2017}. {Following this, the~attackers developed a backdoor in the accountancy software and pushed this out to all users through the update server they had gained control over.} From this vulnerability, the~attackers would deploy the ransomware which would then spread further to other machines on the network by using the EternalBlue exploit. This resulted in large-scale corporate infections, which were only furthered by the lack of SMB version 1's security precautions that should have been implemented not only before, but~also after WannaCry's devastation. 
Although the ransomware initially targeted Ukraine, multinational companies with Ukraine-based offices had caused the ransomware to spread globally. Once a machine became infected, NotPetya would perform the usual ransomware executions steps. However, most notably, NotPetya would not only encrypt files but also alter the Master Boot Record. Once NotPetya had finished execution, it would force the machine to reboot, where the Master Boot Record would then display NotPetya's ransom demands instead of booting the Windows operating system installed. As~a result, domain controllers operating Windows Server would not be able to offer their~services.



 
What makes NotPetya particularly unique is that the encryption process is irreversible, meaning if the ransom is paid the attackers would still be unable to offer the victim their machine's functionality back. At~this point, NotPetya's ransomware classification becomes disputable, as some claim that it is instead a malware classified as a ``wiper'', designed solely to wipe the machine's data, despite the ransom demand present alongside the decryption process~\cite{Mamedov2017,Suiche2017}. The~attack was so devastating that American pharmaceutical company Merck \& Co alone had estimated that by the end of 2017, it had cost them \$870 million in damages, a~number that would then later rise to \$1.3 billion when filing for insurance claims~\cite{Voreacos2019}.

In October 2020, the~United States government would proceed to charge six Russian GRU officers for various cyber-crimes, alleging that one instance of their actions involved the spread of NotPetya in 2017, while the UK government would also expose the other attacks that these people were responsible for~\cite{GovernmentoftheUnitedKingdom2020,Starks2020,UnitedStatesDepartmentofJustice2020}. This was a sentiment expressed as early as 2018, as the governments of the UK, USA, and~Australia had all stated that Russia was behind the NotPetya attack~\cite{Marsh2018,Shaikh2018,Volz2018}. Russia's involvement in this ransomware attack that specifically targeted Ukraine should, then, come as no surprise to those aware of the ongoing Russo-Ukrainian War and should serve as a demonstration of modern civilisation's hybrid warfare~\cite{Limnell2015}. Perhaps NotPetya used the ransom demand to hide their real aim of data destruction, or~the developers wanted to also profit from their attack; however, it is apparent that this was clearly not an ordinary money-focused ransomware. From~the inability to decrypt to the perpetrator's origins, it seems that in this instance, ransomware evolved from the simple profit-chasing malware it started as and became yet another political cyberweapon to add to the~arsenal.

\subsection{Common Tools Used for Ransomware~Analysis}



{One solution that has been widely adopted amongst not just the academic scene, but~also the commercial scene, is the use of virtualisation. Virtualisation involves running software virtually, from~applications to fully emulated machines running operating systems, without~direct access and control of the host machine's hardware. Virtualisation makes use of hypervisors, which allocate physical system resources, such as CPU cores, memory, and~disk space, to~virtual machines~\cite{Sailer2005}. Hypervisors use two different methods to emulate virtual machines. Type 1 hypervisors, known as bare-metal hypervisors, directly run on the host's hardware to house several virtual machines. These are prevalent in the corporate sphere, as large-capacity servers are set up in conjunction with hypervisors such as Microsoft's Hyper-V or VMware's ESXi. Type 2 hypervisors do not run directly from the hardware, and~instead are applications that run from the host's operating system, for~example, VMware Workstation or VirtualBox. What makes virtualisation particularly advantageous, and~almost necessary to perform dynamic malware analysis and research, is the ability to create and manage device snapshots. Snapshots are a copy of a machine's exact state at a certain point in time. When performing malware analysis using virtual machines, a~snapshot can be created before the machine is infected. Once the machine is infected and the research has been conducted, the~machine state can then be reverted quickly and easily to a previous snapshot, all whilst keeping the host machine safe and isolated from the activity performed within the virtual machine. This feature is available with VirtualBox, an~open-source hypervisor developed by Oracle that is available for several different operating systems~\cite{OracleCorporation2021}. Using VirtualBox nearly eliminates the risk involved with handling malware and provides additional benefits that will ease the setup of this experiment.}

Process Monitor is an open-source monitoring software that was initially developed for Windows by Microsoft but is currently also available on Linux. Process Monitor provides thorough information on the system's registry, file system, network, and~process activity and allows for the exporting of recorded activity into a log file for later analysis~\cite{Russinovich2021}. These recorded values are particularly useful for the dynamic analysis of malware, as it encompasses several major system indicators to thoroughly understand how an examined malware sample operates. As~a result, Process Monitor has become commonplace in academic dynamic malware analysis~\cite{Kao2018,Kardile2017,Kendall2007}.

{Since ransomware variants infect and encrypt, often, thousands of files within a single compromised system, the~manual analysis of their impact may be problematic. Hence, a~number of machine learning approaches were developed to detect and analyse this large amount of data specifically within the Windows operating system~\cite{alhawi2018leveraging}. These methods can be further extended through transfer learning approaches to also protect the client machines within a Windows Active Directory~\cite{pires2020convolutional}. However, frequently, machine learning intrusion detection systems are targeted by adversaries that aim to exploit them for their benefit~\cite{papadopoulos2021launching}; hence, these adversarial attacks should be carefully considered by any machine learning algorithm, and further countermeasures should be taken to protect the systems against them~\cite{titcombe2021practical,papadopoulos2021privacy}.}

\subsection{Related~Work}

The topic of ransomware has become a largely researched and discussed topic in the literature, and~the support for Active Directory and Windows Server has been plentiful from both Microsoft and third-party organisations since day one of inception. However, when the two intersect, there is relatively little information available besides the usual basic security measures recommended to organisations. ``Prevention is the best defence'' is a sentiment often echoed when it comes to cybersecurity, with~the FBI even contextually specifying that “Prevention is the most effective defense against ransomware” \cite{FederalBureauofInvestigation2016,Henkes2016,Ragusa2020}. 

WannaCry has been extensively researched due to its popularity, and~has several publications revolving around its analysis, such as the work of~\cite{Kao2018}, 
in which the authors researched WannaCry through both static and dynamic analysis. In~their dynamic analysis, WannaCry's Windows process activity was an area that was briefly explored; however, the~focus was on processes created as a result of WannaCry's execution, and the authors therefore excluded any investigation into the impact of WannaCry on other pre-existing processes~\cite{Kao2018}. As~for network analysis, their focus was on incoming and outgoing communications recorded through Process Hacker and Wireshark; this analysis~was performed on machines running Windows~7.

In 2017, a~Jigsaw-oriented study was conducted with the aim to produce software that detects and halts the execution of ransomware~\cite{Byrne2017}. By~utilising indicators of compromise, the~software would detect the execution process of ransomware and consequently disable network interfaces before shutting down the machine. This would prove very useful to domain environments where ransomware could rapidly spread across the network. The~indicators of change included suspicious mass file activity, which would trigger the software to prevent further file alterations. Related to the organisational impact from this work, their study involved performing a survey of three organisations that were victims of ransomware attacks. The~survey uncovered the preparedness and confidence of small to medium-sized enterprises (SMEs), non-profit organisations, and~credit unions, and discussed the~results.

{Our work differentiates from the previously mentioned studies since it fills the gap currently seen in the literature concerning the impact of ransomware on Windows Server and Active Directory Domain Services environments. In~the following sections, we present our experimental setup that was used to conduct our investigation approach and conclude with valuable findings that were evaluated and discussed extensively.}

\section{Methodology and~Architecture}
\label{sec:methodology}

The aim of this work is to create a suitable testing environment to observe the behavioural patterns of the tested ransomware variants towards networking services. The~desired outcome of the practical investigation is to determine whether a Windows Server service is operational whilst under the control of the tested ransomware variant. However, it is possible that although a service may be responsive, it may not be entirely functional and may be impaired in unpredictable ways. Therefore, in addition to determining whether the service is operational, noting to what extent the service is impacted is also a priority. 
A study regarding the dynamic analysis of WannaCry states that a large part of dynamic analysis involves establishing a baseline environment and comparing differences with that of the infected state~\cite{Kao2018}. This is a critical component of the methodology of our work and is the basis of forming the required results. A~significant technique for dynamic malware analysis is the use of hardware performance counters (HPFs) or other hardware change indicators~\cite{Or-Meir2019}. Although~useful information could be obtained from this method for future work, this method does not suit the purposes of this work, as it would not provide definitive results to determine the functionality of each service; hence, it is out of our scope. A~critical part of any malware analysis research is appropriately handling the malware. Carelessness in handling the malware could cause unwanted harm; therefore, avoiding the use of a live environment is preferable~\cite{Kendall2007}. As~such, virtualisation is adopted as part of the testing strategy through creating and configuring a virtual network and any number of machines. Once the virtual environment is created, a~baseline is then recorded to compare against each of the variables changed throughout the experiment. For~the proposed question, the~baseline is the fully operational server, while the variables are each ransomware variant that is tested throughout the~experiment.


The results obtained are both qualitative and quantitative data in comparison to the baseline, as~alongside the initial ``yes or no'' indication of a service's functionality comes a descriptive breakdown of the extent to which the service is impacted. Both types of results are integrated and presented alongside each other to produce a plethora of information, as~well as provide transparency towards the depth of the results~\cite{Bryman2007}. Quantitative data is useful as it provides an easily digestible, undisputable answer to a question and removes the need for interpretation~\cite{Fujs2019}. {Qualitative data is then also used to provide background to the non-elaborated quantitative results.} It is not enough to determine alone whether a service is functional without elaborating upon how it may be impacted. For~example, if~a network file storage is functional but contains unusable ransomware-encrypted files, then it would be misleading to simply label it as operational. As~such, extra information is~required.

\subsection{Environmental~Controls}

To ensure there is no interference in the execution of ransomware and the functionality of the domain controller's services, an~appropriate configuration must be applied with several measures taken to ensure a sterile virtual environment. Removing any internet connection from both virtual machines and the host machine is necessary for the implementation. This is not only to ensure that the ransomware does not escape the host machine and spread across the host machine's network, but~to also ensure the ransomware does not have contact with any external C\&C servers. This is required for WannaCry, as it attempts to contact the kill switch domain upon start-up, but removing internet access is also used when testing other variants to ensure no external factors impact the~execution.

Typically, when testing the impact of malware, it is recommended to populate the file system with artificial files. Although~this is not a priority in this instance, as the target is the services, artificial files are still used in the case of testing the network file share and webserver. These service's directories are populated by using files that are easily manufactured on the target machine, such as folders, text, and~image files, rather than using files sourced from the internet. To~further ensure no interference from Window's built-in security measures, the~pre-installed Windows Defender anti-malware application should also be disabled on the domain controller. Conversely, Windows Defender should be kept up to date on the host machine to prevent any accidental execution outside of the virtual~environment. 

\subsection{{Services Configuration, Domain Structure and Testing~Strategy}}

{As part of the testing strategy, it must be determined which services are to be tested. Including commonly used services is a priority, with~additional easily configurable services also included for obtaining additional information.  
Regarding the domain structure and logon services, by~logging in from the client machine, the~Key Distribution Center (KDC) service is responsible for user authentication; hence, its functionality is tested. The~KDC service also uses Active Directory's account database to provide its authentication service~\cite{Microsoft2018}.}


{To experimentally test the functionality of the network's file-sharing capabilities, a~home directory can be created on the client user and be accessed externally. Without~using folder redirection, the~default home directory access method does not enable offline caching and accessibility; therefore, no additional group policy objects would need to be configured. 
If the file share remains operational, it is still possible that the ransomware could impact the stored files. An~operational file share with infected files is still valuable information, as~it allows for other security configurations to be implemented by a system administrator, such as file extension change blockers, or~placing the file share on a separate, more frequently updated server.}


{Internet Information Services (IIS) \cite{stanek2007internet}, is Microsoft's proprietary webserver application that is designed to be used with Windows Server. Creating and configuring a web page with IIS on the domain controller would be an optimal service to test the ransomwares' capabilities. As~the webserver is primarily created to be accessed by external machines, it would be an optimal test regarding the purpose of this experiment, as~well as a vulnerable webserver being a likely source for a ransomware infection. Testing this service would involve creating a new HyperText Markup Language (HTML) web page. The~HTML page could then point to potential ransomware target files on the local machine, such as pointing to an image file stored in a common directory, to~further test the extensiveness and the impact of each ransomware variant.}

{Building upon the webserver, the~Domain Name System (DNS) service can also be tested. DNS is used to set up a domain name for the IIS server so that it can be accessed without using the webserver's IP address. The~DNS service intakes requests to resolve the more human-friendly website's domain name to the webserver's IP address~\cite{papadopoulos2020privacy}.
Microsoft suggests that one method of testing whether a target machine is a functioning DNS server is by utilising a PowerShell command~\cite{Microsoft2021a}. The~``Test-DnsServer'' command can be used with the DNS server's IP address as the target parameter and can specify a zone to lookup on the target server. Although~this command can be issued on the target machine itself, ideally this would be done by a different Windows Server machine which would not be set up as part of this investigation.
It would be most beneficial to test these services by accessing the website through the DNS hostname; however, if~DNS does not respond while IIS itself is still functioning, this can still be proven by accessing the IIS server through the server's IP address, thereby bypassing the DNS server that would have otherwise resolved the hostname. If~the webserver is not functioning while DNS is, this can also still be proven by viewing the DNS resolved hostnames cache locally stored on the client machine.}
To test the functionality of the Dynamic Host Configuration Protocol (DHCP) service, the client machine can issue the ``ipconfig /release'' command to remove the automatically assigned IP address. Once the domain controller is infected, an~``ipconfig /renew'' command can be issued to retrieve a new address from the DHCP server. The~client should then receive an IP address within the range specified by the DHCP server if services are unaffected. If~the client fails to receive an address from the server, it would instead obtain one from the Automatic Private IP Addressing feature built into Windows. The~range for these addresses is easily identifiable as it ranges from ``169.254.0.1'' to ``169.254.255.254'' with a subnet mask of ``255.255.0.0''.


Group Policy objects can be easily configured in the Group Policy Management editor. The~domain controller automatically implements a default domain policy that restricts the administrative privileges and rights of standard user accounts. To~test that the group policy service is still functional, an edit can be made to the default domain policy after the client machine has already logged on, as~retrieving group policy is part of the logon process. Once the domain controller is infected, the client machine can issue the ``gpupdate /force'' command to actively retrieve any updates to the group policy since last retrieval. The~change, or~lack thereof, in~group policy would be observed to determine functionality. Ideally, setting a group policy that enforces a specific image file as the Windows desktop background would be a desirable test; however, if~the image file is encrypted and unable to upload, it would be difficult to determine whether the group policy as a whole remains functional. Therefore, an~additional policy is required as a backup test to provide the quantitative data~desired.

\section{Implementation}
\label{sec:implementation}



{The implementation was undertaken on a computer with an Intel quad-core i5-3570K CPU and 16GB of DDR3 RAM. The~machine was formatted, and only the required software was installed to perform the experiments. To~emulate both the domain controller and client machine, VirtualBox was used on the single host machine. VirtualBox was used due to its array of features that would greatly aid in the experiment, as~well as due to its open-source nature. Using VirtualBox for an experiment that involves multiple machines removes the need to physically source the hardware for those machines and allows for much more liberty regarding hardware configurations. Performing the investigation through virtual machines also provides an additional security buffer to prevent sandbox escapism. Perhaps most helpful to the experiment is VirtualBox's snapshot feature that allows a machine's exact state to be saved and later reverted to if necessary. Once the server and client machine were fully configured, a~snapshot was taken of each machine's state. After~the ransomware had been executed and the appropriate results recorded, the~machines were reverted to their uninfected state to start again with the next ransomware variant. VirtualBox provides an additional Guest Additions feature which allows for folders to be shared from the host to the virtual machine~\cite{VirtualBox2021}. This was used to transfer the ransomware executable to the virtual machines that lack internet access. Both the client and server machine were each allocated 1 CPU core with 4GB of RAM, with~50 and 40 GB of hard drive space, respectively.} 


Windows Server 2016 was used for the domain controller, as it is the most recent major revision of the Windows Server operating system, making it more relevant for current and future use interests. Although~Server 2019 superseded 2016, 2019 brought no notable changes that would impact the experiment and is still built upon the same base platform as Server 2016 and Windows 10, making it less popular and likely an unnecessary change that many organisations will opt to skip until Server 2016's end of life. The~client machine operated upon Windows 10 Consumer Edition, which is also the most recent major Windows consumer operating system. The~client machine was fully updated for both Windows Security and Windows Defender updates so that the client machine did not become infected and impact the results from the domain controller responses. 
Realistically, in~an enterprise environment, staff machines would be automatically updated through Windows operating system policy, whilst servers and infrastructure machines would be left operational until unexpected downtime forces user interaction. On~top of this, staff machines require a strong locally installed antivirus to protect the machine from the user, whilst server machines that are open to connection from a multitude of machines would only be hindered by an intrusive antivirus. This results in a more secure client machine paired with an insecure server, therefore justifying cloning this setup for the practical~investigation.

\subsection{Network~Configuration}

\begin{figure}[H]
\includegraphics[width=0.78\linewidth]{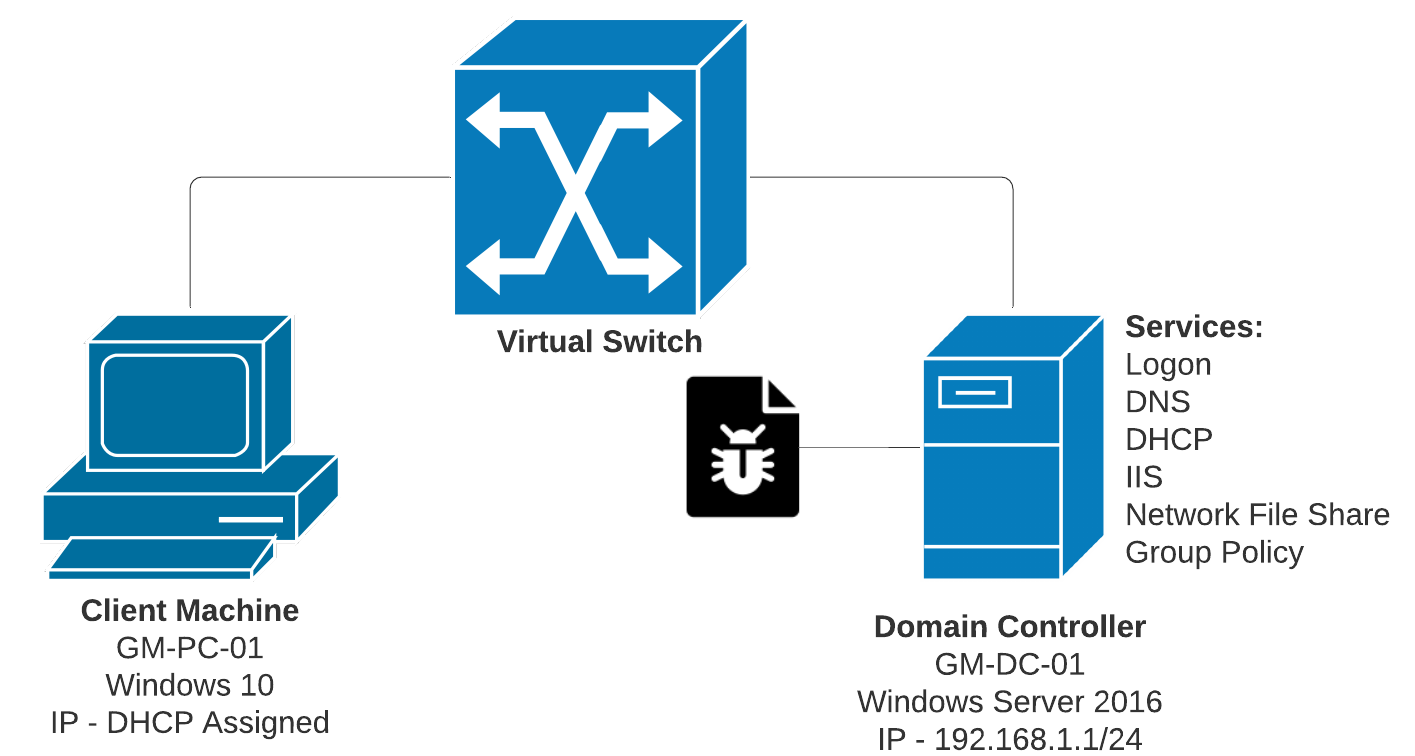}
\caption{Diagram of virtual network~environment. \label{fig6}}
\end{figure}

The virtual machines were configurated to operate within the internal virtual Local Area Network (LAN) environment built into VirtualBox. VirtualBox support drivers automatically create a virtual switch to manage the internal network, then emulate a wired connection from the network interface cards of the virtual machines to the switch~\cite{VirtualBox2021a}. This virtual network environment isolates the virtual machines from the host, as~well as from utilising the host's network adapter for further internet connection. Figure~\ref{fig6} displays the virtual network environment~setup.


The domain controller was assigned a static IP address of 192.168.1.1 with a \linebreak 255.255.255.0 subnet mask. The~client would initially retain an IP address of 192.168.1.2 until it had established membership to the domain. Once the client had connected, it was then set to obtain a dynamic IP address from the DHCP server. Alongside obtaining an IP address, the~client also received the address of the DNS server, which was also that of the domain controller. Table~\ref{tab1} lists the IP addresses relevant to the~experiments.







\begin{table}[H] 
\caption{Network configuration of the~devices.\label{tab1}}
\tablesize{\fontsize{9}{9}\selectfont}
\begin{adjustwidth}{-\extralength}{0cm}
\centering 
\begin{tabular}{m{4.2cm} m{9.9cm} m{3cm}}
\toprule
\textbf{IP Address}	& \textbf{Use Case}  & \textbf{Pertinent Device}\\
\midrule
192.168.1.1     
& Static IP address assigned to the domain controller.		    
& Domain Controller\\\midrule

192.168.1.2     
& Preliminary IP address used by the client machine to connect to the domain.		    
& Client Machine\\\midrule

192.168.1.10-
192.168.1.20 
& IP address range allocated to the DHCP server scope for use by the client machine.	
& Client Machine\\\midrule

169.254.0.1-169.254.255.254
& APIPA range used by the client machine if it is unable to obtain a dynamic IP address from a DHCP server.	
& Client Machine\\

\bottomrule
\end{tabular}
\end{adjustwidth}
\end{table}
\unskip


\subsection{Ransomware Acquisition and~Execution}

\begin{figure}[H]
\includegraphics[width=0.8\linewidth]{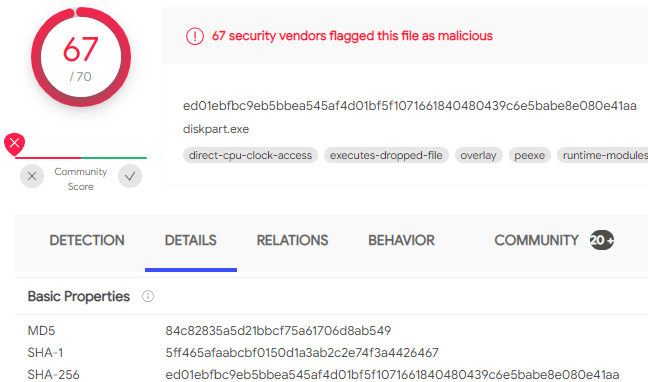}
\caption{Submission 
 of the WannaCry executable to VirusTotal with generated~hashes. \label{fig7}}
\end{figure}

Three ransomware variants were chosen for the investigation. The~WannaCry, TeslaCrypt, and~Jigsaw executables were all acquired from ``theZoo'' GitHub repository~\cite{Ytisf2020}. The~repository was created for the purposes of archiving authentic, easily accessible malware samples for research purposes. The~samples are provided in a zipped, password-protected file to ensure there is no accidental execution. Once the executables were downloaded, they were then uploaded to VirusTotal to verify their authenticity. The submission for WannaCry can be seen in Figure~\ref{fig7}. VirusTotal provides aggregated commercial antivirus and community-based malware identification and produces a cryptographic hash that can be compared to historical publicly documented articles. The~cryptographic hashes generated for WannaCry, TeslaCrypt, and~Jigsaw were all compared against the articles published by FireEye, Secureworks, and~Trend Micro, respectively, to verify their authenticity~\cite{Berry2017,Secureworks2015,TrendMicro2016}.


WannaCry was used due to its lack of sandbox escapism or evasion mechanisms~\cite{Zimba2017}. WannaCry's popularity means that sourcing and verifying an authentic copy is much easier than lesser-known ransomware variants. Older, less contemporary ransomware strains are also more readily available, including Jigsaw and TeslaCrypt, which are available from the same repository as WannaCry. For~testing purposes, the~three ransomware variants' instantaneous execution is more desirable than more sophisticated ransomware strains that lay dormant to perform additional actions such as prior file extraction to a C\&C server for~extortion.

\begin{figure}[H]
\includegraphics[width=0.7\linewidth]{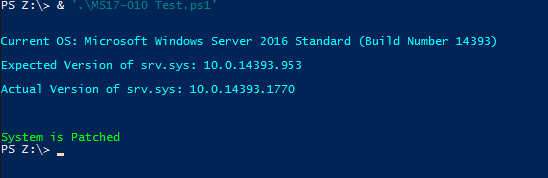}
\caption{Output 
 of PowerShell script stating patch~status. \label{fig8}}
\end{figure}

As part of preparing the domain controller for infection, Windows Defender was removed through the ``Remove Roles and Features Wizard'' in the Server Manager application to ensure no interference in the ransomware's execution. However, the~Server 2016 installation package obtained from Microsoft contained security updates dated to 2018. A~PowerShell script published by Microsoft was then used to determine that the updates installed would provide security against WannaCry and CVE-2017-0144~\cite{Microsoft2021}, with~the output shown in Figure~\ref{fig8}. 


Further investigation showed that the installed security updates were to prevent the SMB vulnerability that WannaCry would exploit for its network-spreading capabilities~\cite{Microsoft2017}. However, as~this exploit only pertains to the network transmission of WannaCry, it does not interfere with the execution process, as that responsibility lies upon Windows Defender, which was not only outdated but also removed from the server. The~transmission method of WannaCry is not relevant to this investigation as the ransomware was placed directly onto the virtual machine from the host; therefore, the~security update can be~ignored. 


\subsection{Additional~Software}

Process Monitor was used throughout testing specifically to record the process activity on the domain controller. The~process action to be monitored for this investigation is the ``process exit'' operation pertaining to the relevant services. After~a successful infection, a~report was produced detailing which processes had terminated. There are few automated report software that would accurately probe each service and perform the desired tasks post-setup, such as testing the additional configured parameters. It is possible that a PowerShell script could be created and utilised on the client machine to probe the required services on the domain controller.

\subsection{Services~Configuration}

Each of the following services were configured to allow for easy probing of the service's functionality, and~where possible, configured with additional parameters that would allow the impact on a typical real-world service~setup to be explored.

\subsubsection{Domain Structure and Logon~Services}

When naming the domain controller, as~seen in Figure~\ref{fig6}, the~hostname ``GM-DC-01'' was applied. It was then used to create the domain ``GM-Domain.com'' to be used for the implementation. Inside Active Directory, a test user account that retained default user access controls was created for ``Christopher Guzman'' with the username ``ChristopherGuzman''. Before~testing, it was ensured that the user was not able to log on using cached credentials which was done by altering a group policy object. In~the default domain policy, the object ``Interactive logon: Number of previous logons to cache'' was defined with a value of zero. This ensures that if the domain controller is unable to process the logon request and authenticate the credentials, the~client is unable to log on. The~client machine was then restarted and attempted to log on with the Christopher Guzman account once the domain controller were~infected. 

\subsubsection{Network File~Share}

Before enabling the home directory for user Christopher Guzman, the~share directory was created. A~folder entitled ``Share'' was created within the root of the C drive. This folder was then shared within the network with a path of ``\textbackslash\textbackslash GM-DC-01\textbackslash Share''. On Christopher's Active Directory account, the~home directory path was specified as the local path of ``C:\textbackslash Share\%USERNAME\%'', where ``\%USERNAME\%''automatically converts to ``ChristopherGuzman''. After~the domain controller had been infected, the~Christopher Guzman account logged onto the client machine and attempted to access the network file share directory. The~state of each file located inside the share directory was also~recorded.


\subsubsection{DNS and IIS Web~Services}

To configure the IIS server, the~default HTML file ``iisstart.html'' stored in ``C:\textbackslash inetpub\textbackslash \linebreak wwwroot'' was replaced with a customised HTML file. The~new HTML file simply contained a text heading, paragraph, and~reference to an image file that was also stored within the wwwroot subdirectory. This file path was also inspected once under infection to observe the impact on the subdirectory. The~client was then used to access the website using the domain name or~IP address as failover, and~the displayed web page contents were noted. As~for DNS, two records were created within the forward lookup zone. The~first was a CNAME record that maps the ``www.gm-site.com
'' alias to the fully qualified domain name of ``GM-DC-01.gm-site.com''. Following this, the~A record was then utilised to point the hostname from the fully qualified domain name to the IP address of the webserver, which in this case remains the same as the domain controller at ``192.168.1.1''. Before~using the client machine to access the webserver after it had been infected, the~command ``ipconfig /flushdns'' was issued on the client machine to clear the DNS cache and force a DNS record retrieval from the DNS server once again. If~IIS were to be unresponsive whilst DNS was still functional, the~``ipconfig /displaydns'' command would be issued to view the cached resolved hostnames acquired from the DNS server. The~browser cache was also cleared to prevent the browser from automatically rendering a non-responsive web page from previously cached files, such as the~image.

\subsubsection{DHCP~Service}

Before configuring the DHCP service for testing, the~client machine was issued a static IP address within the same network as the domain controller to connect to the domain. Once the client machine had connected, the~network adapter was set to obtain an IP address automatically and the machine was then restarted. To~set up the DHCP service for testing, an~IP address range was created. The~configured DHCP scope contained addresses from ``192.168.1.10'' to ``192.168.1.20'' with a subnet mask of ``255.255.255.0''. This removes the conflict from the 192.168.1.1 address held by the domain controller and will help differentiate it from the 192.168.1.2 address used by the client before it had connected to the domain. Once the ``ipconfig /renew'' command had been issued, the~new IP address was noted down and compared to the range set by the DHCP~scope.

\subsubsection{Group~Policy}

Two test policies were created to determine group policy's functionality. 
The first test policy chosen for the experiment was to disable access to the command prompt. By~changing the value of “Prevent access to the command prompt” to enabled, this setting was put into effect. 
This was tested by updating the group policy object on the domain controller, then issuing the ``gpupdate /force'' command on the client machine. Once the group policy had updated, the~command prompt was reopened and checked for the presence of the ``command prompt has been disabled by your administrator'' message, which was observed. 
This test was performed last, as access to the command prompt was needed to flush the DNS cache and test the DHCP service. This method only demonstrates whether the group policy remains operational and does not show how the group policy interacts with files that may be especially vulnerable to ransomware infection. As~a result, a~second test policy was needed. The~second policy that was implemented entailed defining an image file as the default wallpaper. When pushed to the client device, this group policy would cause the client machine to retrieve the image file from the domain controller and set it as the client machine's wallpaper, replacing the default Windows logo. To~do this, an~image file was placed inside a ``wallpaper'' subdirectory of the ``Share'' directory used by the network file share service, and its path was then specified as the target file for the wallpaper~GPO.

\section{Evaluation and~Discussion}
\label{sec:evaluation}



{The results displayed in Table~\ref{tab2} illustrate whether the stated service was fundamentally operational upon ransomware infection of the webserver. All services were functional to a certain degree when tested against all three ransomware variants. This finding was expected; however, the~presented table does not display the extent to which each service was impacted. The~level of impact interference posed by each variant is further elaborated upon in the following subsections.}


{The manual testing method utilised throughout the investigation would not uncover how each service was impacted by its system processes. Process Monitor was also used to verify the results obtained from the manual testing method to examine the processing activity that occurred throughout the ransomware infection. 
The log files produced by Process Monitor showed no alterations to the service processes examined. However, the~log files did produce other notable entries, particularly the ransomware variants terminating the ``vssadmin.exe'' process responsible for volume shadow copy backups. This is a common tactic used by ransomware to prevent the users, and~some commercial backup applications, from~restoring their data through a shadow copy of the disk prior to the ransomware infection~\cite{wecksten2016novel,Nick2018}. By~showing that the ransomware variants terminated this process, we can determine that the ransomware variants executed successfully and did not alter their execution process as a result of sandbox detection.} 
\begin{table}[H] 
\setlength{\tabcolsep}{7.9mm}
\caption{Experimental results. As~expected, all tested services were functional to a degree against all three ransomware~variants. \label{tab2}}
\begin{tabular}{cccc}
\toprule
\textbf{Services}   & \textbf{WannaCry}	& \textbf{TeslaCrypt}	& \textbf{Jigsaw}\\
\midrule
Logon Service & \checkmark & \checkmark & \checkmark \\
Network File Share & \checkmark & \checkmark & \checkmark \\
IIS & \checkmark & \checkmark & \checkmark \\
DNS & \checkmark & \checkmark & \checkmark \\
DHCP & \checkmark & \checkmark & \checkmark \\
Group Policy & \checkmark & \checkmark & \checkmark \\
\bottomrule
\end{tabular}
\end{table}





\subsection{Depth of Impact to~Services}

{Throughout the investigation, several directories and files were examined on the domain controller to ascertain the level of impact inflicted by each ransomware variant. The~files displayed in Table~\ref{tab3} represent files critical to the operation of certain services, as~well as manually placed artificial files for further testing. As~part of the testing strategy, the~commands were utilised either to manually probe the services or to aid in uncovering the extent of infection impact. These were mostly performed on the client device, with~the exception of the ``TestDnsServer'' command, which could not be issued from the client device. In~the following subsections, we extensively present the impact of the ransomware variants on the logon services, network file share, IIS, DNS, DHCP, and~finally, the group policy. Each subsection presents a separate Windows Active Directory service and detailed information within its scope.}


\begin{table}[H] 
\caption{Domain controller files and directories relevant to the investigation~experiments\label{tab3}}

\begin{adjustwidth}{-\extralength}{0cm}
\centering 
\begin{tabular}{m{3cm} m{7cm} m{7.1cm}}
\toprule
\textbf{Domain Service}	& \textbf{Relevant File Path}	& \textbf{Description of Relevant Data}\\
\midrule
Logon Service          
& \path{C:\Windows\NTDS\ntds.dit}		
& Database file containing credentials belonging to Active Directory users\\\midrule

Network File Share     
& \path{C:\Share\ChristopherGuzman\ } and \textbackslash\textbackslash GM-DC-01\textbackslash Share\textbackslash ChristopherGuzman	
& Mapped home directory path, local and network, respectively, for~the user ``Christopher Guzman''\\\midrule

IIS                    
& \path{C:\inetpub\wwwroot\ }		
& Default directory designated to IIS to store files relevant to displaying the web page\\\midrule

DNS                    
& \path{C:\Windows\System32\dns\ }	
& Directory containing DNS records\\\midrule

DHCP                   
& \path{C:\Windows\System32\dhcp\dhcp.mdb}		
& Database file responsible for storing DHCP scope information\\\midrule

Group Policy           
& \path{C:\Share\CommonFiles\wallpaper.png}		
& Path of wallpaper image used for the second test policy\\\midrule
& \path{C:\Windows\SYSVOL\domain\policies} & Default path for storing group policy objects\\
\bottomrule
\end{tabular}
\end{adjustwidth}
\end{table}





\subsubsection{Impact on Logon~Services}

During the investigation experiments, none of the tested ransomware variants caused the client machine to fail the logon test. The~database storing Active Directory credentials used by the logon service (located in the path shown in Table~\ref{tab3}) is a directory that was excluded from WannaCry's encryption scope possibly due to its system-critical nature. This finding shows that not only did the ransomware variants not shut down the relevant service, but~they did not encrypt or impact the database containing user credentials that would be used to log on from the client~machine.

\subsubsection{Impact on Network File~Share}

Throughout all three tests, each ransomware variant produced similar results. The~network file share remained accessible by the client machine; however, all three ransomware variants did encrypt the files stored within the share. Interestingly, the~file path of the network share was a newly created directory on the root of the C drive, showing that the tested ransomware variants did not settle for the standard built-in user directories (such as ``Downloads, Documents, Pictures, etc.'') and instead pushed further by encrypting any user-made directory. This information allows for the exploration of further avenues in terms of file share protection against ransomware, such as using a system files directory, hidden directory, or~mapping to a directory on a different~server. 

\subsubsection{Impact on~IIS}

\begin{figure}[H]
\includegraphics[width=0.4\linewidth]{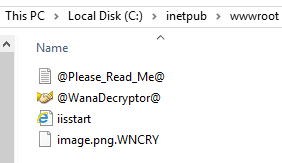}
\caption{IIS 
 root directory under WannaCry's~infection. \label{fig11}}
\end{figure}

\begin{figure}[H]
\includegraphics[width=0.4\linewidth]{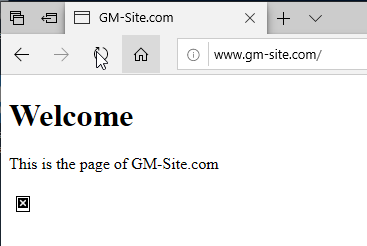}
\caption{Failure to display image referenced by the ``.html''~file. \label{fig12}}
\end{figure}

Furthermore, all ransomware variants allowed the IIS server to stay online but also had similar effects on the IIS root directory. The~first deployment was WannaCry which targeted the ``wwwroot'' directory by dropping its signature text file and application and encrypted pre-existing files in the directory. TeslaCrypt and Jigsaw, on the other hand, only opted to encrypt the files inside the directory. The~image file inside the directory that was referenced by the HTML file was encrypted by all three variants; however, the~HTML file itself was not. This was confirmed for both TeslaCrypt and Jigsaw by selecting their option to output a list of encrypted files, which showed that neither had successfully targeted the HTML file. The~listing of files and their appended extensions in the ``wwwroot'' directory after having been encrypted by WannaCry can be seen in Figure~\ref{fig11}. This resulted in a displayable HTML page that correctly displayed both text sections but did not correctly display the image. In~place of the image was an icon representing an ``X'' as shown in Figure~\ref{fig12}.


\vspace{-12pt}


After acquiring the results and uncovering the failure to encrypt the HTML file from all three variants, it was speculated that, as the IIS service was already running when the ransomware was executed, it may have protected and write-locked the HTML file while it was in use. A~second test with minor alterations disproved this, as~each ransomware variant still opted not to target the HTML file for encryption while the IIS service was stopped. Further research in the literature uncovered the full list of file extensions targeted by the three variants tested and concluded that all three excluded the ``.html'' extension from their target scope~\cite{Abrams2016,Grinler2015,SecureworksCounterThreatUnitResearchTeam2017}.

\subsubsection{Impact on~DNS}

\begin{figure}[H]
\includegraphics[width=0.4\linewidth]{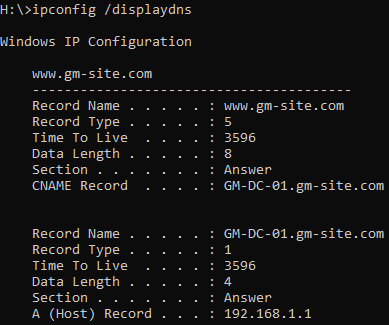}
\caption{DNS record displayed on the client~machine. \label{fig13}}
\end{figure}

As seen in Table~\ref{tab2}, the~DNS service remained operational alongside the infection of all three variants. As~IIS was operational, the~web page responded to the client machine that accessed the web page using the ``www.gm-site.com'' URL, eliminating the need to test the IIS service using the server IP address. Using the ``displaydns'' command parameter on the client machine stated in Table~\ref{tab4} also showed that the DNS server provided a full, correct record, as seen in Figure~\ref{fig13}. Moreover, a~PowerShell command to test the DNS service was utilised to test if the target server IP represented a functional DNS server. There is little room for interference with the DNS service due to the method of storing DNS-centric data. The~DNS records are all stored inside a system-critical ``system32'' subdirectory and appended with a ``.dns'' file extension~\cite{Liu2003}; therefore, it would be extremely unusual for a ransomware variant to target the DNS records themselves, even through a blanket encryption strategy, unless~it was manufactured specifically to target a server~environment.

\begin{table}[H] 
\caption{Executed commands in the investigation~experiments.\label{tab4}}
\begin{adjustwidth}{-\extralength}{0cm}
\centering 
\begin{tabular}{m{4cm} m{3cm} m{10.1cm}}
\toprule
\textbf{Command}	& \textbf{Device Used}	& \textbf{Command Description}\\
\midrule
ipconfig /release                       & Client		    
& Removes the DHCP allocated IP address assigned to the network interface\\\midrule
ipconfig /renew                         & Client		    
& Retrieves an IP address to assign from the DHCP server\\\midrule
ipconfig /flushdns                      & Client		    
& Clears the system's cached DNS records \\\midrule
ipconfig /displaydns                    & Client		    
& Displays the cached DNS records\\\midrule
gpupdate /force                         & Client		    
& Forces the machine to update all group policy settings\\\midrule
Test-DnsServer (with IP Address 192.168.1.1)   & Domain Controller	
& PowerShell command to test whether the target IP address is an operational DNS server\\
\bottomrule
\end{tabular}
\end{adjustwidth}
\end{table}
\unskip

\subsubsection{Impact on~DHCP}

Similarly to DNS, the~DHCP service is difficult to interfere with, outside of outright stopping the service, which neither three variants managed to do. The~DHCP service also stores its files inside of a subdirectory of ``system32'' and utilises no other files from standard consumer-friendly directories. The~client machine showed no issue with obtaining an IP address from the DHCP server by using the appropriate commands from all three variants. The~DHCP server manager clearly displayed the live IP release and renewal as the client machine issued the respective commands, which could be seen in the DHCP server manager's application GUI, as this was also left operational by all three ransomware~variants.

\subsubsection{Impact on Group~Policy}

Unsurprisingly, group policy also remained functional with similar interruptions to the tested area of the service. The~first test involved utilising a policy that would disable access to the command prompt for a standard user account, which proved successful when updating the policy on the client machine whilst the domain controller was infected (file paths shown in Table~\ref{tab3}). 
The second test that set the default wallpaper for use by the client machine involved defining the path of the image file used as a wallpaper. This pointed to the file inside the ``Share'' directory that was targeted by all three variants and, as~a result, the~image file was encrypted. The~test resulted in the client machine failing to apply the policy and replacing the default Windows logo wallpaper image with an empty, black wallpaper. This demonstrates the group policy's ability to stay operational during the infection; however, it also reveals the inability to protect and hide relevant additional files to the~service.

\section{Conclusions}
\label{sec:conclusion}

{The primary focus of this work was to produce information regarding ransomware and its impact on Windows Server environments for use by organisations and enterprises. Since our investigation activities were performed post-infection from the ransomware variants, there is no computational overhead to the infrastructure upon its regular operation.
The hypothesis stated that ransomware would not stop the tested services but rather impact their functionality through alternative means, such as encrypting pertinent files. 
Our implementation involved creating a virtual environment with a domain controller operating Windows Server 2016 and a client machine running Windows 10. Several Windows Server services tested were then configured to allow for extensive testing with the intent to produce qualitative and quantitative data for results. 
From the three tested ransomware variants, all tested services remained operational. The~services that utilised files not belonging to the service's default configurations and file paths did see interruptions to their functionality, whilst system-critical paths remained untouched. This proved the previously stated hypothesis true.}

Although the tested services remained operational, these were all proprietary Microsoft services offered within the Windows Server package that crypto-ransomware variants could wrongly identify as system-critical rather than system processes that could be stopped without repercussions. In~a future study, testing third party applications from computer-oriented software to the software responsible for physical entities could produce vastly different results, as third-party software does not typically use system-critical file paths. Using the physical office key card entry system example, testing this would also not be possible with the current physical hardware limitations, and~with a larger budget, more commercial software and products could be tested, providing further valuable information to organisations. 
Moreover, Windows Server offers a wide array of built-in services that could see different results when tried against this testing method. Additionally, third-party applications can also be run on Windows Server machines, particularly in corporate domain environments, that do not retain the same operating system protections and priorities as built-in packages. However, throughout this experiment, the~most critical variable was the ransomware variants used. These could provide unique alterations in results and would be the next step to uncovering the full extent of ransomware on domain~services.

\vspace{6pt}

\authorcontributions{All authors contributed to the conceptualization and methodology of the manuscript; G.M. performed data preparation and the practical experiments; G.M., P.P. and N.P. contributed in writing; J.A. and W.J.B. reviewed and edited the manuscript. All authors have read and agreed to the published version of the~manuscript.}

\funding{This 
 research received no external~funding.}

\institutionalreview{Not applicable.}

\informedconsent{Not applicable.}

\dataavailability{Data are contained within the article.} 

\conflictsofinterest{The authors declare no conflict of~interest.}

\begin{adjustwidth}{-\extralength}{0cm}
\reftitle{References}

\end{adjustwidth}


\begin{thebibliography}{999}

\bibitem[Franke(2017)]{Franke2017}
Franke, U.
\newblock {The cyber insurance market in Sweden}.
\newblock {\em Comput. Secur.} {\bf 2017}, {\em 68},~130--144. https://doi.org/10.1016/j.cose.2017.04.010.

\bibitem[{Datto Inc.}(2020)]{DattoInc.2020}
{Datto Inc.}
\newblock {\emph{Ransomware Report}}; 2020. 
Available online: \url{https://www.datto.com/resources/dattos-global-state-of-the-channel-ransomware-report} (accessed on 25 Jan 2022).


\bibitem[Huang \em{et~al.}(2018)Huang, Aliapoulios, Li, Invernizzi, Bursztein,
  McRoberts, Levin, Levchenko, Snoeren, and McCoy]{Huang2018}
Huang, D.Y.; Aliapoulios, M.M.; Li, V.G.; Invernizzi, L.; Bursztein, E.;
  McRoberts, K.; Levin, J.; Levchenko, K.; Snoeren, A.C.; McCoy, D.
\newblock {Tracking Ransomware End-to-end}.
\newblock  In {Proceedings of the }2018 39th IEEE Symposium on Security and Privacy (SP), 21-23 May 2018, San Francisco, CA, USA; 2018;
 pp.
  618--631. https://doi.org/10.1109/SP.2018.00047.

\bibitem[Ramsey(2017)]{Ramsey2017}
Ramsey, D.
\newblock {\emph{Google, UC San Diego and NYU Estimate \$25 Million in Ransomware  Payouts}}; 2017. Available online: \url{https://ucsdnews.ucsd.edu/pressrelease/google_uc_san_diego_and_nyu_estimate_25_million_in_ransomware_payouts} (accessed on 25 Jan 2022).

\bibitem[Sophos(2019)]{Sophos2019}
Sophos.
\newblock {\emph{Sophos Labs 2019 Threat Report}}; 2019. Available online: \url{https://www.sophos.com/en-us/medialibrary/PDFs/technical-papers/sophoslabs-2019-threat-report.pdf} (accessed on 25 Jan 2022).

\bibitem[Sophos(2018)]{Sophos2018}
Sophos.
\newblock {\emph{SamSam: The (Almost) Six Million Dollar Ransomware}}; 2018. Available online: \url{https://www.sophos.com/en-us/medialibrary/PDFs/technical-papers/SamSam-The-Almost-Six-Million-Dollar-Ransomware.pdf} (accessed on 25 Jan 2022).

\bibitem[{Deep Instinct}(2020)]{DeepInstinct2020}
{Deep Instinct}.
\newblock {\emph{Cyber Threat Landscape Report 2019--2020}}; 2020. Available online: \url{https://info.deepinstinct.com/hubfs/Cyber_Threat_Landscape_Report_2019-2020.pdf} (accessed on 25 Jan 2022).

\bibitem[{Help Net Security}(2014)]{HelpNetSecurity2014}
{Help Net Security}.
\newblock {\emph{Active Directory Flaw Impacts 95\% of Fortune 1000 Companies}}; 2014. Available online: \url{https://www.helpnetsecurity.com/2014/07/15/active-directory-flaw-impacts-95-of-fortune-1000-companies/} (accessed on 25 Jan 2022).

\bibitem[Lexico(2020)]{Lexico2020}
Lexico.
\newblock {\emph{Domain|Definition of Domain by Oxford Dictionary on Lexico.com}}; 2020. Available online: \url{https://www.lexico.com/definition/domain} (accessed on 25 Jan 2022).

\bibitem[Microsoft(2020)]{Microsoft2020a}
Microsoft.
\newblock {\emph{How to Detect, Enable and Disable SMBv1, SMBv2, and SMBv3 in  Windows}}; 2020. Available online: \url{https://docs.microsoft.com/en-us/windows-server/storage/file-server/troubleshoot/detect-enable-and-disable-smbv1-v2-v3} (accessed on 25 Jan 2022).

\bibitem[Alspach(2019)]{Alspach2019}
Alspach, K.
\newblock {\emph{Microsoft Inspire 2019: The 6 Biggest Statements from Gavriella  Schuster And Judson Althoff}}; 2019. Available online: \url{https://www.crn.com/news/channel-programs/microsoft-inspire-2019-the-6-biggest-statements-from-gavriella-schuster-and-judson-althoff} (accessed on 25 Jan 2022).

\bibitem[Sayer(2020)]{Sayer2020}
Sayer, P.
\newblock {\emph{Not Dead Yet: Windows Server 2008 Users Have Options}}; 2020. Available online: \url{https://www.cio.com/article/3514735/not-dead-yet-windows-server-2008-users-have-options.html} (accessed on 25 Jan 2022).

\bibitem[Lexico(2020)]{Lexico2020a}
Lexico.
\newblock {\emph{Malware|Definition of Malware by Oxford Dictionary on Lexico.com}};
   2020. Available online: \url{https://www.lexico.com/definition/malware} (accessed on 25 Jan 2022).
   

\bibitem[Liska and Gallo(2016)]{Liska2016}
Liska, A.; Gallo, T.
\newblock {\em {Ransomware: Defending against Digital Extortion}}; O'Reilly Media Inc.: ISBN: 1491967854; 
  2016. Available online: \url{https://books.google.com.ec/books?id=IIORDQAAQBAJ} (accessed on 25 Jan 2022).

\bibitem[Kaspersky(2020)]{Kaspersky2020}
Kaspersky.
\newblock {\emph{What Are the Different Types of Ransomware?}}; 2020. Available online: \url{https://www.kaspersky.co.uk/resource-center/threats/ransomware-examples} (accessed on 25 Jan 2022).

\bibitem[Brewer(2016)]{Brewer2016}
Brewer, R.
\newblock {Ransomware attacks: Detection, prevention and cure}.
\newblock {\em Netw. Secur.} {\bf 2016}, {\em 2016},~5--9. https://doi.org/10.1016/S1353-4858(16)30086-1.

\bibitem[Adamov and Carlsson(2017)]{Adamov2017}
Adamov, A.; Carlsson, A.
\newblock {The state of ransomware. Trends and mitigation techniques}.
\newblock  In Proceedings of the 2017 IEEE East-West Design \& Test Symposium (EWDTS), 2017; 29 September - 2 October 2017, Novi Sad, Serbia; pp.
  1--8. https://doi.org/10.1109/EWDTS.2017.8110056.

\bibitem[Subedi \em{et~al.}(2018)Subedi, Budhathoki, and Dasgupta]{Subedi2018}
Subedi, K.P.; Budhathoki, D.R.; Dasgupta, D.
\newblock {Forensic Analysis of Ransomware Families Using Static and Dynamic
  Analysis}.
\newblock  In Proceedings of the 2018 IEEE Security and Privacy Workshops (SPW), 2018; May 24 2018, San Francisco, CA, USA; pp. 180--185. https://doi.org/10.1109/SPW.2018.00033.

\bibitem[Hassan(2019)]{Hassan2019}
Hassan, N.A.
\newblock {\emph{Ransomware Families BT---Ransomware Revealed: A Beginner's Guide to  Protecting and Recovering from Ransomware Attacks}}; Apress: Berkeley, CA, USA,   2019; pp. 47--68. https://doi.org/10.1007/978-1-4842-4255-1\_3.

\bibitem[Zimba and Chishimba(2019)]{Zimba2019}
Zimba, A.; Chishimba, M.
\newblock {On the Economic Impact of Crypto-ransomware Attacks: The State of
  the Art on Enterprise Systems}.
\newblock {\em Eur. J. Secur. Res.} {\bf 2019}, {\em
  4},~3--31. https://doi.org/10.1007/s41125-019-00039-8.

\bibitem[Cartwright and Cartwright(2019)]{Cartwright2019}
Cartwright, A.; Cartwright, E.
\newblock {Ransomware and reputation}.
\newblock {\em Games} {\bf 2019}, {\em 10},~26. https://doi.org/10.3390/g10020026.

\bibitem[Coveware(2020)]{Coveware2020a}
Coveware.
\newblock {\emph{Ransomware Costs Double in Q4 as Ryuk, Sodinokibi Proliferate}};
  2020. Available online: \url{https://www.coveware.com/blog/2020/1/22/ransomware-costs-double-in-q4-as-ryuk-sodinokibi-proliferate} (accessed on 25 Jan 2022). 

\bibitem[Abrams(2016)]{Abrams2016a}
Abrams, L.
\newblock {\emph{UltraCrypter Not Providing Decryption Keys after Payment. Launches
  Help Desk}}; 2016. Available online: \url{https://www.bleepingcomputer.com/news/security/ultracrypter-not-providing-decryption-keys-after-payment-launches-help-desk/} (accessed on 25 Jan 2022).

\bibitem[Sophos(2020)]{Sophos2020}
Sophos.
\newblock {\emph{The State of Ransomware 2020}}; 2020. Available online: \url{https://www.sophos.com/en-us/medialibrary/Gated-Assets/white-papers/sophos-the-state-of-ransomware-2020-wp.pdf} (accessed on 25 Jan 2022).

\bibitem[Meland \em{et~al.}(2020)Meland, Bayoumy, and Sindre]{Meland2020}
Meland, P.H.; Bayoumy, Y.F.F.; Sindre, G.
\newblock {The Ransomware-as-a-Service economy within the darknet}.
\newblock {\em Comput. Secur.} {\bf 2020}, {\em 92},~101762. https://doi.org/https://doi.org/10.1016/j.cose.2020.101762.

\bibitem[Hernandez-Castro \em{et~al.}(2017)Hernandez-Castro, Cartwright, and
  Stepanova]{Hernandez-Castro2017}
Hernandez-Castro, J.; Cartwright, E.; Stepanova, A.
\newblock {Economic analysis of ransomware}.
\newblock {\em arXiv preprint arXiv:1703.06660}; 2017. 

\bibitem[{National Cyber Security
  Centre}(2020)]{NationalCyberSecurityCentre2020}
{National Cyber Security Centre}.
\newblock {\emph{The NCSC Annual Review 2020}}; 2020. Available online: \url{https://www.ncsc.gov.uk/files/Annual-Review-2020.pdf} (accessed on 25 Jan 2022).


%
%

\bibitem[Papadopoulos \em{et~al.}(2021)Papadopoulos, Pitropakis, and
  Buchanan]{papadopoulos2021decentralised}
Papadopoulos, P.; Pitropakis, N.; Buchanan, W.J., Decentralised Privacy: A
  Distributed Ledger Approach.
\newblock In {\em Handbook of Smart Materials, Technologies, and Devices:  Applications of Industry 4.0}; Hussain, C.M., Di~Sia, P., Eds.; Springer  International Publishing: Cham,  Switzerland, 2021; pp. 1--26. https://doi.org/10.1007/978-3-030-58675-1\_58-1.

\bibitem[Bistarelli \em{et~al.}(2018)Bistarelli, Parroccini, and
  Santini]{Bistarelli2018}
Bistarelli, S.; Parroccini, M.; Santini, F.
\newblock {Visualizing Bitcoin Flows of Ransomware: WannaCry One Week Later.}
\newblock {InProceedings of the Italian Conference on Cybersecurity (ITASEC), Milan, Italy, 6--9 February 2018}.

\bibitem[Kshetri and Voas(2017)]{Kshetri2017}
Kshetri, N.; Voas, J.
\newblock {Do Crypto-Currencies Fuel Ransomware?}
\newblock {\em IT Prof.} {\bf 2017}, {\em 19},~11--15. https://doi.org/10.1109/MITP.2017.3680961.

\bibitem[Young \em{et~al.}(2021)Young, Chrysoulas, Pitropakis, Papadopoulos,
  and Buchanan]{young2021evaluating}
Young, E.H.; Chrysoulas, C.; Pitropakis, N.; Papadopoulos, P.; Buchanan, W.J.
\newblock Evaluating Tooling and Methodology when Analysing Bitcoin Mixing
  Services After Forensic Seizure.
\newblock  In Proceedings of the IEEE 2021 International Conference on Data Analytics for Business and  Industry (ICDABI), 26-27 October 2020, Sakheer, Bahrain. 2021; pp. 650--654.

\bibitem[Lemmou and Souidi(2018)]{Lemmou2018}
Lemmou, Y.; Souidi, E.M.
\newblock {Infection, Self-reproduction and Overinfection in Ransomware: The
  Case of TeslaCrypt}.
\newblock  In Proceedings of the 2018 International Conference on Cyber Security and Protection of  Digital Services (Cyber Security), 11-12 June 2018, Glasgow, Scotland, United Kingdom. 2018; pp. 1--8. https://doi.org/10.1109/CyberSecPODS.2018.8560670.

\bibitem[Mimoso(2016)]{Mimoso2016}
Mimoso, M.
\newblock {\emph{Decryption Utilities Unlock Files Encrypted by All TeslaCrypt  Versions}}; 2016. Available online: \url{https://threatpost.com/decryption-utilities-unlock-files-encrypted-by-all-teslacrypt-versions/118602/} (accessed on 25 Jan 2022).



\bibitem[Villeneuve(2015)]{Villeneuve2015}
Villeneuve, N.
\newblock {\emph{TeslaCrypt: Following the Money Trail and Learning the Human Costs  of Ransomware}}; 2015. Available online: \url{https://www.fireeye.com/blog/threat-research/2015/05/teslacrypt_followin.html} (accessed on 25 Jan 2022).

\bibitem[O'Kane \em{et~al.}(2018)O'Kane, Sezer, and Carlin]{OKane2018}
O'Kane, P.; Sezer, S.; Carlin, D.
\newblock {Evolution of ransomware}.
\newblock {\em IET Netw.} {\bf 2018}, {\em 7},~321--327.

\bibitem[Conti \em{et~al.}(2018)Conti, Gangwal, and Ruj]{Conti2018}
Conti, M.; Gangwal, A.; Ruj, S.
\newblock {On the economic significance of ransomware campaigns: A Bitcoin
  transactions perspective}.
\newblock {\em Comput. Secur.} {\bf 2018}, {\em 79},~162--189.

\bibitem[Europol(2017)]{Europol2017}
Europol.
\newblock {\emph{Over 28,000 Devices Decrypted and 100+ Global Partners---No More  Ransom Celebrates Its First Year}}; 2017. Available online: \url{https://www.europol.europa.eu/newsroom/news/over-28-000-devices-decrypted-and-100-global-partners-–-no-more-ransom-celebrates-its-first-year} (accessed on 25 Jan 2022).

\bibitem[{National Cyber Security
  Centre}(2017)]{NationalCyberSecurityCentre2017}
{National Cyber Security Centre}.
\newblock {\emph{Weekly Threat Report 22nd December 2017}}; 2017. Available online: \url{https://www.ncsc.gov.uk/report/weekly-threat-report-22nd-december-2017} (accessed on 25 Jan 2022).

\bibitem[Microsoft(2018)]{Microsoft2018a}
Microsoft.
\newblock {\emph{Microsoft SMB Protocol and CIFS Protocol Overview}}; 2018. Available online: \url{https://docs.microsoft.com/en-us/windows/win32/fileio/microsoft-smb-protocol-and-cifs-protocol-overview} (accessed on 25 Jan 2022).

\bibitem[{Revert Service}(2020)]{RevertService2020}
{Revert Service}.
\newblock \emph{Server (LanmanServer) Service Defaults in Windows 10}; 2020. Available online: \url{http://revertservice.com/10/lanmanserver/} (accessed on 25 Jan 2022).

\bibitem[Microsoft(2020{\natexlab{a}})]{Microsoft2020}
Microsoft.
\newblock \emph{SMBv1 Is Not Installed by Default in Windows 10 Version 1709,  Windows Server Version 1709 and Later Versions}; 2020. Available online: \url{https://docs.microsoft.com/en-us/windows-server/storage/file-server/troubleshoot/detect-enable-and-disable-smbv1-v2-v3} (accessed on 25 Jan 2022).

\bibitem[Microsoft(2020{\natexlab{b}})]{Microsoft2020b}
Microsoft.
\newblock \emph{Overview---Product End of Support}; 2020. Available online: \url{https://docs.microsoft.com/en-us/lifecycle/overview/product-end-of-support-overview} (accessed on 25 Jan 2022).

\bibitem[Microsoft(2017)]{Microsoft2017}
Microsoft.
\newblock \emph{Microsoft Security Bulletin MS17-010---Critical}; 2017. Available online: \url{https://docs.microsoft.com/en-us/security-updates/securitybulletins/2017/ms17-010} (accessed on 25 Jan 2022).

\bibitem[{The MITRE Corporation}(2017)]{TheMITRECorporation2017}
{The MITRE Corporation}.
\newblock \emph{CVE-2017-0144}; 2017. Available online: \url{https://cve.mitre.org/cgi-bin/cvename.cgi?name=CVE-2017-0144} (accessed on 25 Jan 2022).

\bibitem[MalwareTech(2017)]{MalwareTech2017}
MalwareTech.
\newblock \emph{How to Accidentally Stop a Global Cyber Attacks}; 2017. Available online: \url{https://www.malwaretech.com/2017/05/how-to-accidentally-stop-a-global-cyber-attacks.html} (accessed on 25 Jan 2022).

\bibitem[Suiche(2017)]{Suiche2017a}
Suiche, M.
\newblock \emph{WannaCry—Decrypting files with WanaKiwi + Demos}; 2017. Available online: \url{https://blog.comae.io/wannacry-decrypting-files-with-wanakiwi-demo-86bafb81112d} (accessed on 25 Jan 2022).

\bibitem[Castillo and Falzon(2018)]{Castillo2018}
Castillo, D.; Falzon, J.
\newblock {An analysis of the impact of Wannacry cyberattack on cybersecurity  stock returns}.
\newblock {\em Rev. Econ. Financ.} {\bf 2018}, {\em 13},~93--100.

\bibitem[SentinelOne(2019)]{SentinelOne2019}
SentinelOne.
\newblock \emph{Eternalblue|The NSA-Developed Exploit That Just Won't Die}; 2019. Available online: \url{https://www.sentinelone.com/blog/eternalblue-nsa-developed-exploit-just-wont-die/} (accessed on 25 Jan 2022).

\bibitem[Whittaker(2019)]{Whittaker2019}
Whittaker, Z.
\newblock \emph{Two Years after WannaCry, a Million Computers Remain at Risk}; 2019. Available online: \url{https://techcrunch.com/2019/05/12/wannacry-two-years-on/} (accessed on 25 Jan 2022).
  

\bibitem[Doyle-Price(2019)]{Doyle-Price2019}
Doyle-Price, J.
\newblock \emph{NHS: Computer Software}; 2019. Available online: \url{https://questions-statements.parliament.uk/written-questions/detail/2019-07-10/275828} (accessed on 25 Jan 2022).

\bibitem[Ghafur \em{et~al.}(2019)Ghafur, Kristensen, Honeyford, Martin, Darzi,
  and Aylin]{Ghafur2019}
Ghafur, S.; Kristensen, S.; Honeyford, K.; Martin, G.; Darzi, A.; Aylin, P.
\newblock {A retrospective impact analysis of the WannaCry cyberattack on the
  NHS}.
\newblock {\em NPJ Digit. Med.} {\bf 2019}, {\em 2},~1--7.

\bibitem[Stamatellis \em{et~al.}(2020)Stamatellis, Papadopoulos, Pitropakis,
Katsikas, and Buchanan]{stamatellis2020privacy}
Stamatellis, C.; Papadopoulos, P.; Pitropakis, N.; Katsikas, S.; Buchanan, W.J.
\newblock A Privacy-Preserving Healthcare Framework Using Hyperledger Fabric.
\newblock {\em Sensors} {\bf 2020}, {\em 20},~6587.

\bibitem[{Skybox Security}(2018)]{SkyboxSecurity2018}
{Skybox Security}.
\newblock \emph{TSMC WannaCry Hits OT Plants with a Hefty Price Tag}; 2018. Available online: \url{https://www.skyboxsecurity.com/blog/tsmc-wannacry/} (accessed on 25 Jan 2022).

\bibitem[Goud(2018)]{Goud2018}
Goud, N.
\newblock \emph{Details about WannaCry Ransomware Attack on Boeing Company}; 2018. Available online: \url{https://www.cybersecurity-insiders.com/details-about-wannacry-ransomware-attack-on-boeing-company/} (accessed on 25 Jan 2022).

\bibitem[Muncaster(2018)]{Muncaster2018}
Muncaster, P.
\newblock {Boeing Computers Hit by WannaCry}.
\newblock {\em Infosecurity Mag.} {\bf 2018}. 
Available online: \url{https://www.infosecurity-magazine.com/news/boeing-computers-hit-by-wannacry/} (accessed on 25 Jan 2022).

\bibitem[Mackenzie(2019)]{Mackenzie2019}
Mackenzie, P.
\newblock \emph{The WannaCry Hangover}; 2019. Available online: \url{https://news.sophos.com/en-us/2019/09/18/the-wannacry-hangover/} (accessed on 25 Jan 2022).

\bibitem[{British Broadcasting
  Corporation}(2017)]{BritishBroadcastingCorporation2017}
{British Broadcasting Corporation}.
\newblock \emph{Ukraine Cyber-Attack: Software Firm MeDoc's Servers Seized}; 2017. Available online: \url{https://www.bbc.co.uk/news/technology-40497026} (accessed on 25 Jan 2022).

\bibitem[Mamedov and Ivanov(2017)]{Mamedov2017}
Mamedov, O.; Ivanov, A.
\newblock \emph{ExPetr/Petya/NotPetya is a Wiper, Not Ransomware}; 2017. Available online: \url{https://securelist.com/expetrpetyanotpetya-is-a-wiper-not-ransomware/78902/} (accessed on 25 Jan 2022).

\bibitem[Suiche(2017)]{Suiche2017}
Suiche, M.
\newblock \emph{Petya.2017 Is a Wiper Not a Ransomware}; 2017. Available online: \url{https://blog.comae.io/petya-2017-is-a-wiper-not-a-ransomware-9ea1d8961d3b} (accessed on 25 Jan 2022).

\bibitem[Voreacos \em{et~al.}(2019)Voreacos, Chiglinsky, and
  Griffin]{Voreacos2019}
Voreacos, D.; Chiglinsky, K.; Griffin, R.
\newblock \emph{Merck Cyberattack's \$1.3 Billion Question: Was It an Act of War?}; 2019. Available online: \url{https://www.bloomberg.com/news/features/2019-12-03/merck-cyberattack-s-1-3-billion-question-was-it-an-act-of-war} (accessed on 25 Jan 2022).

\bibitem[{Government of the United
  Kingdom}(2020)]{GovernmentoftheUnitedKingdom2020}
{Government of the United Kingdom}.
\newblock \emph{UK Exposes Series of Russian Cyber Attacks against Olympic and  Paralympic Games}; 2020. Available online: \url{https://www.gov.uk/government/news/uk-exposes-series-of-russian-cyber-attacks-against-olympic-and-paralympic-games} (accessed on 25 Jan 2022).

\bibitem[Starks(2020)]{Starks2020}
Starks, T.
\newblock \emph{US Charges Russian GRU Officers for NotPetya, Other Major Hacks}; 2020. Available online: \url{https://www.cyberscoop.com/russian-hackers-notpetya-charges-gru/} (accessed on 25 Jan 2022).

\bibitem[{United States Department of
  Justice}(2020)]{UnitedStatesDepartmentofJustice2020}
{United States Department of Justice}.
\newblock \emph{Six Russian GRU Officers Charged in Connection with Worldwide  Deployment of Destructive Malware and Other Disruptive Actions in  Cyberspace}; 2020. Available online: \url{https://www.justice.gov/opa/pr/six-russian-gru-officers-charged-connection-worldwide-deployment-destructive-malware-and} (accessed on 25 Jan 2022).


\bibitem[Marsh(2018)]{Marsh2018}
Marsh, S.
\newblock \emph{US Joins UK in Blaming Russia for NotPetya Cyber-Attack}; 2018. Available online: \url{https://www.theguardian.com/technology/2018/feb/15/uk-blames-russia-notpetya-cyber-attack-ukraine} (accessed on 25 Jan 2022).

\bibitem[Shaikh(2018)]{Shaikh2018}
Shaikh, R.
\newblock \emph{US, UK, Australia Warn Russia of “International Consequences”---NotPetya Outbreak Attributed to the Kremlin}; 2018. Available online: \url{https://wccftech.com/australia-us-uk-russia-notpetya/} (accessed on 25 Jan 2022).

\bibitem[Volz and Young(2018)]{Volz2018}
Volz, D.; Young, S.
\newblock \emph{White House Blames Russia for 'Reckless' NotPetya Cyber Attack}; 2018. Available online: \url{https://www.reuters.com/article/us-britain-russia-cyber/uk-blames-russia-for-cyber-attack-moscow-decries-western-campaign-idUSKCN1FZ0Q3} (accessed on 25 Jan 2022).

\bibitem[Limn{\'{e}}ll(2015)]{Limnell2015}
Limn{\'{e}}ll, J.
\newblock {The exploitation of cyber domain as part of warfare: Russo-Ukrainian
  war}.
\newblock {\em Int. J. Cyber-Secur. Digit. Forensics}
  {\bf 2015}, {\em 4},~521--533.

\bibitem[Sailer \em{et~al.}(2005)Sailer, Valdez, Jaeger, Perez, {Van Doorn},
  Griffin, Berger, Sailer, Valdez, and Jaeger]{Sailer2005}
Sailer, R.; Valdez, E.; Jaeger, T.; Perez, R.; {Van Doorn}, L.; Griffin, J.L.;
  Berger, S.; Sailer, R.; Valdez, E.; Jaeger, T.
\newblock {sHype: Secure hypervisor approach to trusted virtualized systems}.
\newblock {\em Tech. Rep. RC23511}; 2005. 
Available online: \url{https://www.paramecium.org:4443/~leendert/publications/rc23511.pdf} (accessed on 25 Jan 2022).

\bibitem[{Oracle Corporation}(2021)]{OracleCorporation2021}
{Oracle Corporation}.
\newblock \emph{Oracle VM VirtualBox}; 2021. Available online: \url{https://www.virtualbox.org/} (accessed on 25 Jan 2022).


\bibitem[Russinovich(2021)]{Russinovich2021}
Russinovich, M.
\newblock \emph{Process Monitor v3.61}; 2021. Available online: \url{https://docs.microsoft.com/en-us/sysinternals/downloads/procmon} (accessed on 25 Jan 2022).

\bibitem[Kao and Hsiao(2018)]{Kao2018}
Kao, D.; Hsiao, S.
\newblock {The dynamic analysis of WannaCry ransomware}.
\newblock In Proceedings of the 2018 20th International Conference on Advanced Communication
  Technology (ICACT), Chuncheon-si, 11 - 14 February 2018, Gangwon-do, South Korea; 2018; pp. 159--166. https://doi.org/10.23919/ICACT.2018.8323682.

\bibitem[Kardile(2017)]{Kardile2017}
Kardile, A.B.
\newblock \emph{Crypto Ransomware Analysis and Detection Using Process Monitor}; 2017. Available online: \url{https://rc.library.uta.edu/uta-ir/handle/10106/27184} (accessed on 25 Jan 2022).

\bibitem[Kendall and McMillan(2007)]{Kendall2007}
Kendall, K.; McMillan, C.
\newblock {Practical malware analysis}.
\newblock  In Proceedings of the Black Hat Conference, USA, 2007; p.~10. Available online: \url{https://www.blackhat.com/presentations/bh-dc-07/Kendall_McMillan/Paper/bh-dc-07-Kendall_McMillan-WP.pdf} (accessed on 25 Jan 2022).

\bibitem[Alhawi \em{et~al.}(2018)Alhawi, Baldwin, and
  Dehghantanha]{alhawi2018leveraging}
Alhawi, O.M.; Baldwin, J.; Dehghantanha, A.
\newblock Leveraging machine learning techniques for windows ransomware network
  traffic detection. In {\em Cyber Threat Intelligence}; Springer: Berlin/Heidelberg, Germany, 
  2018; pp.
  93--106.

\bibitem[Pires~de Lima and Marfurt(2020)]{pires2020convolutional}
Pires~de Lima, R.; Marfurt, K.
\newblock Convolutional neural network for remote-sensing scene classification:
  Transfer learning analysis.
\newblock {\em Remote Sens.} {\bf 2020}, {\em 12},~86.

\bibitem[Papadopoulos \em{et~al.}(2021)Papadopoulos, Essen, Pitropakis,
  Chrysoulas, Mylonas, and Buchanan]{papadopoulos2021launching}
Papadopoulos, P.; Essen, O.T.V.; Pitropakis, N.; Chrysoulas, C.; Mylonas, A.;
  Buchanan, W.J.
\newblock Launching Adversarial Attacks against Network Intrusion Detection
  Systems for IoT.
\newblock {\em J. Cybersecur. Priv.} {\bf 2021}, {\em
  1},~252--273.

\bibitem[Titcombe \em{et~al.}(2021)Titcombe, Hall, Papadopoulos, and
  Romanini]{titcombe2021practical}
Titcombe, T.; Hall, A.J.; Papadopoulos, P.; Romanini, D.
\newblock Practical Defences Against Model Inversion Attacks for Split Neural
  Networks.
\newblock {\em arXiv} {\bf 2021}, arXiv:2104.05743.

\bibitem[Papadopoulos \em{et~al.}(2021)Papadopoulos, Abramson, Hall, Pitropakis, and Buchanan]{papadopoulos2021privacy}
Papadopoulos, P.; Abramson, W.; Hall, A.J.; Pitropakis, N.; Buchanan, W.J.
\newblock Privacy and Trust Redefined in Federated Machine Learning.
\newblock {\em Mach. Learn. Knowl. Extr.} {\bf 2021}, {\em 3}, 333-356.

\bibitem[{Federal Bureau of
  Investigation}(2016)]{FederalBureauofInvestigation2016}
{Federal Bureau of Investigation}.
\newblock {\emph{Ransomware Prevention and Response for CISOs—FBI}}; 2016. Available online: \url{https://www.fbi.gov/file-repository/ransomware-prevention-and-response-for-cisos.pdf/view} (accessed on 25 Jan 2022).

\bibitem[Henkes(2016)]{Henkes2016}
Henkes, A.
\newblock \emph{Prevention Is the Best Defense: Five Key Measures to Stop Malware}; 2016. Available online: \url{https://www.cybered.io/webinars/prevention-best-defense-five-key-measures-to-stop-malware-w-1104} (accessed on 25 Jan 2022).

\bibitem[Ragusa(2020)]{Ragusa2020}
Ragusa, J.
\newblock \emph{CyberSecurity: Advice for Prevention}; 2020. Available online: \url{https://www.metropolitanrisk.com/cyber-security-prevention-is-the-best-defense/} (accessed on 25 Jan 2022).

\bibitem[Byrne and Thorpe(2017)]{Byrne2017}
Byrne, D.; Thorpe, C.
\newblock {Jigsaw: An investigation and countermeasure for ransomware attacks}.
\newblock In \emph{European Conference on Cyber Warfare and Security}; Academic Conferences International Limited, 29-30 June 2017, Dublin, Ireland; 
  2017; pp. 656--665. 

\bibitem[Or-Meir \em{et~al.}(2019)Or-Meir, Nissim, Elovici, and
  Rokach]{Or-Meir2019}
Or-Meir, O.; Nissim, N.; Elovici, Y.; Rokach, L.
\newblock {Dynamic malware analysis in the modern era—A state of the art
  survey}.
\newblock {\em ACM Comput. Surv.} {\bf 2019}, {\em 52}, 1--48. https://doi.org/10.1145/3329786.

\bibitem[Bryman(2007)]{Bryman2007}
Bryman, A.
\newblock {Barriers to Integrating Quantitative and Qualitative Research}.
\newblock {\em J. Mixed Methods Res.} {\bf 2007}, {\em 1},~8--22. https://doi.org/10.1177/2345678906290531.

\bibitem[Fujs \em{et~al.}(2019)Fujs, Miheli{\v{c}}, and Vrhovec]{Fujs2019}
Fujs, D.; Miheli{\v{c}}, A.; Vrhovec, S.L.R.
\newblock {The power of interpretation: Qualitative methods in cybersecurity
  research}.
\newblock In  Proceedings of the 14th International Conference on Availability,  Reliability and Security, 26-29 August 2019, New York, NY, United States; 2019; pp. 1--10.

\bibitem[Microsoft(2018)]{Microsoft2018}
Microsoft.
\newblock \emph{Key Distribution Center}; 2018. Available online: \url{https://docs.microsoft.com/en-us/windows/win32/secauthn/key-distribution-center} (accessed on 25 Jan 2022).

\bibitem[Stanek(2007)]{stanek2007internet}
Stanek, W.
\newblock {\em Internet Information Services (IIS) 7.0 Administrator's Pocket  Consultant}; Microsoft Press; 2007. Available online: \url{https://www.oreilly.com/library/view/internet-information-services/9780735623644/} (accessed on 25 Jan 2022).

\bibitem[Papadopoulos \em{et~al.}(2020)Papadopoulos, Pitropakis, Buchanan, Lo,
  and Katsikas]{papadopoulos2020privacy}
Papadopoulos, P.; Pitropakis, N.; Buchanan, W.J.; Lo, O.; Katsikas, S.
\newblock Privacy-Preserving Passive DNS.
\newblock {\em Computers} {\bf 2020}, {\em 9},~64.

\bibitem[Microsoft(2021)]{Microsoft2021a}
Microsoft.
\newblock \emph{How to Verify That MS17-010 Is Installed}; 2021. Available online: \url{https://support.microsoft.com/en-us/topic/how-to-verify-that-ms17-010-is-installed-f55d3f13-7a9c-688c-260b-477d0ec9f2c8} (accessed on 25 Jan 2022).

\bibitem[VirtualBox(2021{\natexlab{a}})]{VirtualBox2021}
VirtualBox.
\newblock \emph{Chapter 4. Guest Additions}; 2021. Available online: \url{https://www.virtualbox.org/manual/ch04.html} (accessed on 25 Jan 2022).

\bibitem[VirtualBox(2021{\natexlab{b}})]{VirtualBox2021a}
VirtualBox.
\newblock \emph{Chapter 6. Virtual Networking}; 2021. Available online: \url{https://www.virtualbox.org/manual/ch06.html} (accessed on 25 Jan 2022).

\bibitem[Ytisf(2020)]{Ytisf2020}
Ytisf.
\newblock \emph{theZoo/Malwares/Binaries at Master {\textperiodcentered}
  ytisf/theZoo {\textperiodcentered} GitHub}; 2020. Available online: \url{https://github.com/ytisf/theZoo/tree/master/malwares/Binaries} (accessed on 25 Jan 2022).

\bibitem[Berry \em{et~al.}(2017)Berry, Homan, and Eitzman]{Berry2017}
Berry, A.; Homan, J.; Eitzman, R.
\newblock \emph{WannaCry Malware Profile}; 2017. Available online: \url{https://www.fireeye.com/blog/threat-research/2017/05/wannacry-malware-profile.html} (accessed on 25 Jan 2022).

\bibitem[Secureworks(2015)]{Secureworks2015}
Secureworks.
\newblock \emph{TeslaCrypt Ransomware}; 2015. Available online: \url{https://www.secureworks.com/research/teslacrypt-ransomware-threat-analysis} (accessed on 25 Jan 2022).

\bibitem[{Trend Micro}(2016)]{TrendMicro2016}
{Trend Micro}.
\newblock \emph{JIGSAW Crypto-Ransomware Turns Customer-Centric, Uses Chat for  Ransom Attempts}; 2016. Available online: \url{https://blog.trendmicro.com/trendlabs-security-intelligence/jigsaw-crypto-ransomware-turns-customer-centric-uses-chat-ransom-attempts/} (accessed on 25 Jan 2022).

\bibitem[Zimba \em{et~al.}(2017)Zimba, Simukonda, and Chishimba]{Zimba2017}
Zimba, A.; Simukonda, L.; Chishimba, M.
\newblock {Demystifying Ransomware Attacks: Reverse Engineering and Dynamic
  Malware Analysis of WannaCry for Network and Information Security}.
\newblock {\em Zamb. ICT J.} {\bf 2017}, {\em 1},~35--40. https://doi.org/10.33260/zictjournal.v1i1.19.

\bibitem[Microsoft(2021)]{Microsoft2021}
Microsoft.
\newblock \emph{Test-DnsServer (DnsServer)|Microsoft Docs}; 2021. Available online: \url{https://support.microsoft.com/en-us/topic/how-to-verify-that-ms17-010-is-installed-f55d3f13-7a9c-688c-260b-477d0ec9f2c8} (accessed on 25 Jan 2022).

\bibitem[Weckst{\'e}n \em{et~al.}(2016)Weckst{\'e}n, Frick, Sj{\"o}str{\"o}m,
  and J{\"a}rpe]{wecksten2016novel}
Weckst{\'e}n, M.; Frick, J.; Sj{\"o}str{\"o}m, A.; J{\"a}rpe, E.
\newblock A novel method for recovery from Crypto Ransomware infections.
\newblock  In Proceedings of the 2016 2nd IEEE International Conference on Computer and
  Communications (ICCC), 14-17 October 2016, Chengdu, China; 2016; pp. 1354--1358.

\bibitem[{Nick d7xTech}(2018)]{Nick2018}
{Nick d7xTech}.
\newblock \emph{CryptoPrevent, Ransomware Threat Mitigation, and VSSAdmin.exe}; 2018. Available online: \url{https://www.d7xtech.com/cryptoprevent-ransomware-threat-mitigation-and-vssadmin-exe/} (accessed on 25 Jan 2022).

\bibitem[Abrams(2016)]{Abrams2016}
Abrams, L.
\newblock {Jigsaw Ransomware Decrypted: Will Delete Your Files Until You Pay  the Ransom}; 2016. Available online: \url{https://www.bleepingcomputer.com/news/security/jigsaw-ransomware-decrypted-will-delete-your-files-until-you-pay-the-ransom/} (accessed on 25 Jan 2022).

\bibitem[Grinler(2015)]{Grinler2015}
Grinler.
\newblock \emph{New TeslaCrypt Ransomware Sets Its Scope on Video Gamers}; 2015. Available online: \url{https://www.bleepingcomputer.com/forums/t/568525/new-teslacrypt-ransomware-sets-its-scope-on-video-gamers/} (accessed on 25 Jan 2022).

\bibitem[{Secureworks Counter Threat Unit Research
  Team}(2017)]{SecureworksCounterThreatUnitResearchTeam2017}
{Secureworks Counter Threat Unit Research Team}.
\newblock \emph{WCry Ransomware Analysis}; 2017. Available online: \url{https://www.secureworks.com/research/wcry-ransomware-analysis} (accessed on 25 Jan 2022).

\bibitem[Liu \em{et~al.}(2003)Liu, Larson, and Allen]{Liu2003}
Liu, C.; Larson, M.; Allen, R.
\newblock {\em {DNS on Windows Server 2003: Mastering the Domain Name System}};
  O'Reilly Media Inc.: ISBN: 0-596-00562-8; 2003.

\end{thebibliography}
\end{document}